\documentclass [final,onecolumn,notitlepage,a4paper,12pt]{article}
\usepackage [a4paper] {geometry}
\usepackage{amssymb}
\usepackage{amsfonts}
\usepackage{amsmath}
\usepackage[dvips]{graphicx}
\usepackage{hyperref}

\numberwithin{equation}{section}

\newcommand{\R}{\mathbb{R}}
\newcommand{\C}{\mathbb{C}}
\newcommand{\Z}{\mathbb{Z}}
\newcommand{\I}{\mathbb{I}}

\newcommand{\CA}{{\cal A}}

\newcommand{\CC}{{\cal C}}

\newcommand{\CF}{{\cal F}}

\newcommand{\CL}{{\cal L}}

\newcommand{\CN}{{\cal N}}

\renewcommand{\b}{\beta}
\newcommand{\g}{\gamma}
\newcommand{\eps}{\epsilon}

\renewcommand{\i}{\iota}
\renewcommand{\k}{\kappa}

\newcommand{\s}{\sigma}
\newcommand{\w}{\omega}

\renewcommand{\(}{\left(}
\renewcommand{\)}{\right)}
\renewcommand{\[}{\left[}
\renewcommand{\]}{\right]}

\newcommand{\pd}{\partial}

\renewcommand{\exp}[1]{{e}^{#1}}
\newcommand{\ul}{\underline}
\renewcommand{\slash}[1]{/\!\!\!\! #1}

\newcommand{\sign}{{\rm sign}}

\renewcommand{\Im}[1]{\,\mathcal{I}\mathrm{m}\!\left[#1\right]}
\renewcommand{\Re}[1]{\,\mathcal{R}\mathrm{e}\!\left[#1\right]}

\begin{document}
\begin{titlepage}
\begin{center}

{\hbox to\hsize{\hfill WIS/02/08-JAN-DPP}}

\vspace{3cm}

{\LARGE On the Conformal Field Theory Duals of type IIA\\[2mm]
  $AdS_4$ Flux Compactifications}\\[1.5cm]

{\large Ofer Aharony$^{}$,
  Yaron E. Antebi$^{}$,
  Micha Berkooz$^{}$
}\\[8mm]

{\it Department of Particle Physics,\\ Weizmann Institute of
  Science, Rehovot 76100, Israel}

\medskip

{\tt E-mails : Ofer.Aharony@weizmann.ac.il, ayaron@weizmann.ac.il, Micha.Berkooz@weizmann.ac.il}

\vspace*{1.5cm}

\begin{abstract}
\noindent
  We study the conformal field theory dual of the type IIA flux compactification
  model of DeWolfe, Giryavets,
  Kachru and Taylor, with all moduli stabilized.  We find its central
  charge and properties of its operator spectrum.  We concentrate on
  the moduli space of the conformal field theory, which we investigate through domain walls in the
  type IIA string theory.  The moduli space turns out to consist of many
  different branches. We use Bezout's theorem and Bernstein's theorem to
  enumerate the different branches of the moduli space and estimate
  their dimension.
\end{abstract}

\end{center}
\end{titlepage}
\newpage

\section{Introduction and Summary of Results}
\label{sec:introduction}

Flux compactifications of string theory (for reviews see
\cite{Frey:2003tf,Silverstein:2004id,Grana:2005jc,Polchinski:2006gy,Douglas:2006es,Denef:2007pq}) populate large parts of the string landscape,
and may describe our universe. However, the theoretical basis for the
construction of these compactifications is still far from rigorous
(see \cite{Banks:2004xh} for criticism), and is based on using
low-energy supergravity actions in a regime which is different from
the flat-space regime where they are usually derived from string theory. It would be very
interesting if a non-perturbative construction of some flux
compactifications could be found, providing further support for their
consistency, and perhaps leading to new methods for their analysis.  A
promising arena for such a construction is in flux compactifications
involving four dimensional anti-de Sitter (AdS) space. Such
compactifications are dual, by the AdS/CFT correspondence
\cite{Maldacena:1997re,Witten:1998qj,Gubser:1998bc}, to three
dimensional conformal field theories.  Thus, understanding the three
dimensional conformal field theory dual to some $AdS_4$ flux
compactification would give a non-perturbative definition for that
background. Eventually we would like to study the statistics of
conformal field theories that are dual to $AdS_4$ backgrounds, in
order to learn about statistics of flux compactifications, and to try
to understand how to describe also backgrounds with a positive
cosmological constant.

For general flux compactifications, it seems that understanding the
dual conformal field theory must be very complicated (see
\cite{Fabinger:2003gp,Silverstein:2003jp} for attempts in this direction). This is because the
cosmological constant of the resulting background, which is related to
the central charge of the dual conformal field theory, depends in a
very complicated way on the fluxes, and it seems that extremely complicated
dynamics is needed to reproduce this on the field theory side. The
situation seems to be much simpler in the type IIA flux
compactifications constructed in \cite{DeWolfe:2005uu} (these solutions were
further analyzed from the ten dimensional point of view in
\cite{Lust:2004ig,Acharya:2006ne}). These
backgrounds have a ``large-flux limit'' in which some of the fluxes
(the three four-form fluxes $f_4^1$, $f_4^2$ and $f_4^3$) are taken to
be large, such that in that limit the cosmological constant becomes
small, the string coupling becomes weak, and the compact space becomes
large. This means that these backgrounds can reliably be studied
in the supergravity approximation (except perhaps near the orientifold
where the string coupling may be large), and that in the ``large-flux
limit'' the properties of the dual conformal field theories depend in
a simple way on the fluxes.  One can then hope to reproduce this
simple dependence in some field theoretic model. A first attempt at such
an analysis, in a different ``large-flux limit'' which does not lead
to a weakly coupled string theory (not all four-form fluxes are taken
to be large) appeared in \cite{Banks:2006hg}; we will attempt here to
describe the field theories appearing in the generic ``large-flux
limit'', which is described by a weakly coupled string theory.

The naive way to construct a field theory dual for flux backgrounds is
to imagine constructing the flux gradually from branes carrying that
flux, in a manner similar to that in which the $AdS_5\times S^5$
background of string theory is constructed from D3-branes in flat
space. In particular, the 4-form fluxes $f_4^i$ in our background are
carried by D4-branes wrapped on 2-cycles in the compact space, and it
is natural to imagine building the background from such D4-branes
\cite{Fabinger:2003gp,Silverstein:2003jp,Kounnas:2007dd}. It is certainly possible to go from a
background with a large flux (which is already a weakly coupled weakly
curved background) to a background with an even larger flux by adding
such branes, and we will use this in our discussion of the moduli
space of the conformal field theory.  However, it is not clear if one
can construct the full theory from such branes, since in the limit of
a small flux the background becomes not only strongly curved (this
happens also for D3-branes) but also strongly coupled. Nevertheless,
it is still natural to guess that the dual conformal field theory
arises from some decoupled low-energy theory living on three sets of
D4-branes. However, we will find that assuming that the degrees of freedom in this
theory are weakly coupled open strings (in adjoint and bi-fundamental
representations of the resulting $U(f_4^1)\times U(f_4^2)\times
U(f_4^3)$ gauge theory) leads to a contradiction, since the central
charge of the dual conformal field theory (which scales as
$(f_4^1 f_4^2 f_3^3)^{3/2}$, as we will compute in
section \ref{sec:dual-cft}) grows faster than the number of such
degrees of freedom. Thus, the field theory must be more complicated
than the naive theory of open strings, perhaps involving a larger gauge group,
or
\cite{Silverstein:2003jp} fields in multi-fundamental or other
higher representations, or perhaps not coming from any gauge theory at
all.

In order to find clues about this mysterious field theory we
investigate in some detail its moduli space, which can be described
using configurations of domain walls in $AdS_4$. Of course, generic
flux backgrounds preserve no supersymmetry so they would not be
expected to have a moduli space. The flux backgrounds of
\cite{DeWolfe:2005uu} preserve a four dimensional $\CN=1$
supersymmetry, so they are dual to three dimensional $\CN=1$
superconformal field theories. This amount of supersymmetry is not
enough to protect the moduli space from quantum corrections, since
generic scalar potentials are consistent with three dimensional
$\CN=1$ supersymmetry. Nevertheless, in our study (performed in the
weak coupling weak curvature limit) we will find a large moduli space
in these backgrounds. We expect this moduli space to be lifted by
quantum corrections (perhaps non-perturbative), but these quantum
corrections are small in the ``large flux limit'', and we expect the
existence of a moduli space in this limit to be a useful clue for the
construction of the dual field theory. The moduli space turns out to
be very complicated, with many different branches that may be
interconnected. For each such branch we employ some mathematical
theorems that count the number of solutions of polynomial equations,
in order to compute its dimension. We will show that for
large values of the fluxes, the dimension of the moduli space scales as
$\sum_{i < j} f_4^if_4^j$.

The effective field theory at generic points on the moduli space
includes $U(1)$ gauge fields, scalars and fermions; however, in $2+1$
dimensions a $U(1)$ gauge field is equivalent to a compact scalar, so
the presence of these gauge fields does not necessarily imply that the
full theory is related to a $U(f_4^1)\times U(f_4^2)\times U(f_4^3)$
gauge theory. However, there are special submanifolds of the moduli space
in which we can see gauge groups corresponding to all subgroups of
$U(f_4^1)\times U(f_4^2)\times U(f_4^3)$, suggesting that the conformal
field theory may be described as the low-energy limit of some gauge theory
which includes this gauge group. This is further
supported by the scaling of the dimension of the moduli space, that is reminiscent of
strings in the bi-fundamental representation of each pair of gauge groups (and
such bi-fundamental fields indeed appear on the special submanifolds
mentioned above).

So far we have not been able to find a simple field theory model that
would reproduce all the properties that we find; in particular it seems
hard to explain the large number of degrees of freedom, and the
complicated form of the moduli space. We hope that these
properties will provide useful clues for the construction of such a
field theory in the future.

We begin in section \ref{sec:supergravity-model} with a review of the
type IIA backgrounds of \cite{DeWolfe:2005uu} that we will be studying
and of their supersymmetry equations. In section \ref{sec:dual-cft} we
compute various basic properties of the dual field theory, like its
central charges and the generic features of its operator spectrum. In
section \ref{sec:domain-walls} we consider branes spanning
domain walls in the $AdS_4$ space, and find the condition that they
preserve supersymmetry.
We then go on in section \ref{sec:geom-moduli-space} to study the
structure of their moduli space. We compute the moduli space
explicitly for a simple example and find some properties, such as the
dimension, for the generic case. In the appendices we include some
additional calculations, including an explicit calculation of the
supersymmetry in the bulk in appendix \ref{sec:supersymm-equat-bulk}.
In appendix \ref{sec:bps-condition} we show that the domain walls
found in section \ref{sec:domain-walls} obey the BPS condition, and in
appendix \ref{sec:other-domain-walls} we consider the possibility of 
additional domain wall brane configurations.

\section{The Model}
\label{sec:supergravity-model}

In this section we review the low-energy limit of the
background of massive type IIA string theory described by an orientifold of type
IIA string theory on $T^6/\Z_3^2$. This model was studied extensively
in \cite{DeWolfe:2005uu}, where it was shown that by turning on generic
values for the background fluxes it is possible to stabilize all
moduli without the use of non-perturbative effects.
We will start by reviewing the geometrical properties of the compact
manifold, and then discuss the possible moduli and the way in which
they can be stabilized. Finally we will show that the background satisfies the supersymmetry
equations in the bulk.

\subsection{The Geometry}
\label{sec:geometry}

The compact space is an orbifold of $T^6$. We parameterize the torus by
the three complex coordinates $z_i=x_i+iy_i$, with $i=1,2,3$. We
take the complex structure moduli of the tori to be
$\tau_i=\alpha\equiv e^{2\pi i/6}$, so that the $z_i$ coordinates are
periodic with the identifications
\begin{equation}
  \label{eq:1}
  z_i \simeq z_i+1 \simeq z_i+\alpha.
\end{equation}
At this point in the moduli space of the torus, the $T^6$ has a $\Z_3$
symmetry, under which the coordinates transform as
\begin{equation}
  \label{eq:2}
  z_i\to \alpha^2 z_i.
\end{equation}
It is then possible to orbifold by this symmetry. This gives rise to a
singular space, with 27 singular points corresponding to the fixed
points of the $\Z_3$ symmetry \cite{Dixon:1985jw,Strominger:1985ku}.
After this identification, there is a second $\Z_3$ symmetry acting
freely on the coordinates as
\begin{equation}
  \label{eq:3}
  (z_1,z_2,z_3)\to
  (\alpha^2 z_1+\frac{1+\alpha}{3},
  \alpha^4 z_2+\frac{1+\alpha}{3},
  z_3+\frac{1+\alpha}{3}).
\end{equation}
This symmetry identifies triplets of fixed points, thus leading,
after a second orbifold by the second $\Z_3$ symmetry,
to a singular Calabi-Yau manifold with only 9 singular
points (that can be locally described as a $C^3/\Z_3$ singularity). The
cohomology of this manifold is given by $h^{2,1}=0$ and $h^{1,1}=12$.  There
are therefore no complex structure moduli and 12 K\"ahler moduli.
Nine of them are associated to blow-up modes of the singular points, while the
other three K\"ahler moduli describe the volume of the three tori. These
volume moduli $\gamma_i$ appear in the metric as
\begin{equation}
  \label{eq:4}
  ds^2=\sum_{i=1}^3\gamma_i dz^i d\bar z^i,
\end{equation}
or in the K\"ahler form for the manifold as
\begin{equation}
  \label{eq:5}
  J=ig_{i\bar j} \ dz^i\wedge d\bar z^j
  =\sum_{i=1}^3 i\frac{\g_i}{2} dz^i\wedge d\bar z^i.
\end{equation}

It will be useful to write an explicit basis for the cohomology of the compact
space.  There are no one-forms, since the two
$\Z_3$ orbifolds project out all of the one-forms of the torus. There are
three two-forms that form the basis of the untwisted part of $H^2$.
These are the two-forms that remain invariant under the
$\Z_3^2$, and they can be chosen as
\begin{equation}
  \label{eq:6}
  w_i=(\kappa\sqrt3)^{1/3}idz^i\wedge d\bar z^i,
\end{equation}
in an arbitrary normalization (in which the triple intersection is
$\kappa$).  Their Poincar\'e-dual four-forms form the basis for the untwisted
part of $H^4$,
\begin{equation}
  \label{eq:7}
  \tilde w^i=\(\frac3\kappa\)^{1/3} (idz^j\wedge d\bar z^j)\wedge(idz^k\wedge d\bar z^k),
\end{equation}
where $\{i,j,k\}$ are different elements of the set $\{1,2,3\}$. We
choose the normalizations such that
\begin{equation}
  \label{eq:8}
  \int_{T^6/\Z_3^2}w_1\wedge w_2\wedge w_3=\kappa,\qquad
  \int_{T^6/\Z_3^2}w_i\wedge \tilde w^j=\delta_i^j.
\end{equation}
There are also two-forms and four-forms associated with the blow-up modes of
the orbifold fixed points, which we will not write down explicitly.

Since $h^{2,1}=0$, the only 3-forms in the compact geometry are the
holomorphic 3-form
\begin{equation}
  \label{eq:9}
  \Omega=\sqrt{\g_1\g_2\g_3}i dz_1\wedge dz_2\wedge dz_3
\end{equation}
and its complex conjugate $\bar \Omega$.  These are normalized such
that
\begin{equation}
  \label{eq:10}
  \frac i8 \int_{T^6/\Z_3^2} \Omega\wedge\bar\Omega=vol(T^6/\Z_3^2)=\frac{1}{8\sqrt3}\g_1\g_2\g_3,
\end{equation}
and can be verified to obey the standard relations
\begin{equation}
  \label{eq:11}
  J\wedge\Omega=0,\quad
  \frac{i}{8}\Omega\wedge\bar\Omega=\frac{1}{3!}J^3.
\end{equation}

As a last step in defining the geometry we quotient by an
orientifold action. We will use the orientifold of $T^6/\Z_3^2$
presented in \cite{Blumenhagen:1999ev}. The total orientifold action
is given by $\Omega(-1)^{F_L}\sigma$, where $\Omega$ is reflection on
the worldsheet, $F_L$ is the worldsheet left moving fermion number, and
$\sigma$ is the spacetime involution
\begin{equation}
  \label{eq:12}
  z_i\to-\bar z_i.
\end{equation}
Under this action there is a 3 dimensional space left fixed, given by
$\Re{z_i}=0$. Thus, the theory contains an $O6$-plane wrapping this
3-cycle and filling the non compact directions.

Under the orientifold action the different forms have non trivial
transformation properties. The forms defined above transform as
\begin{equation}
  \label{eq:13}
  w_i\to -w_i,\qquad
  \tilde w^i \to \tilde w^i,\qquad
  \Omega\to\bar\Omega.
\end{equation}
One can write the three-forms in a diagonal basis with respect to the
orientifold by decomposing $\Omega$ to its real and imaginary parts,
$\Omega=\frac{\sqrt{\gamma_1\gamma_2\gamma_3}}{3^{1/4}\sqrt2}
(\alpha_0+i\beta_0)$. These transform as
\begin{eqnarray}
  \label{eq:14}
  \alpha_0\to \alpha_0,\qquad \beta_0\to-\beta_0.
\end{eqnarray}

\subsection{Moduli and Their Stabilization}
\label{sec:moduli-their-stab}

In order to stabilize all the moduli we will need to turn on a 10-form
(or 0-form) RR flux, so that in the low-energy limit we obtain Romans' massive IIA
supergravity theory
\cite{Romans:1985tz}  (with a mass parameter proportional to the RR 0-form
field strength), compactified to four dimensions on the $T^6/\Z_3^2$
orientifold discussed in the previous subsection. In addition to the
background metric and dilaton, the theory includes a NS-NS 2-form
$B_2$ (whose field strength is $H_3$), and a RR 1-form and
3-form, $C_1$ and $C_3$ (with field strengths $F_2$ and $F_4$).
\footnote{Note that we use
  the following conventions for the RR fields. We follow the convention
  of \cite{DeWolfe:2005uu,Grimm:2004ua} including an additional factor of $\sqrt{2}$
  with respect to the standard convention, while working with signs as
  in \cite{Martucci:2005ht}. So, we use opposite signs for $F_0$ and $F_6$
  compared to \cite{DeWolfe:2005uu,Grimm:2004ua}.}


Before turning on fluxes, the massless spectrum includes the K\"ahler
parameters from the metric, $\gamma_i$, and the dilaton $\phi$.
Since $B_2$ is odd under $\Omega$, its zero modes are related to the
forms $\w_i$ in the $\sigma$-odd cohomology $H^{2}_-$, and it can be
expanded as
\begin{equation}
  \label{eq:15}
  B_2=\sum b_i \w^i.
\end{equation}
The three zero modes $b_i$ combine with $\gamma_i$ to form the bosonic
part of a chiral multiplet.
Similarly we can expand the RR forms. Since $h^1=0$, the one-form has no
zero modes. The three-form, being even under $\Omega$, has one zero mode,
related to the unique even three-form, $\alpha_0$. Thus we have
\begin{equation}
  \label{eq:16}
  C_3=\xi\alpha_0.
\end{equation}
The four dimensional axiodilaton superfield contains the combination of this
axion $\xi$ with the dilaton $\phi$.


All of these moduli can be stabilized by turning on fluxes along the compact
directions. In order to preserve Poincar\'e invariance, the fluxes can
be written as
\begin{equation}
  \label{eq:17}
  F_n=\hat F_n+vol_4\wedge\tilde F_{n-4},
\end{equation}
where all the indices in $\hat F$ and $\tilde F$ are internal, and they
are Poincar\'e dual using the 6 dimensional metric, $\tilde
F_n=(-1)^{(n-1)(n-2)/2}*_6\hat F_{6-n}$.  The background values for the
fluxes can then be written by expanding the fields in the relevant
cohomology (having the correct parity under the orientifold) :
\begin{eqnarray}
  \label{eq:18}
  H_3=-p\beta_0,\quad
  \hat F_0=-m_0,\quad
  \hat F_2=-m_i w_i,\quad
  \hat F_4=e_i\tilde w^i,\quad
  \hat F_6=-e_0\frac{\alpha_0\wedge\beta_0}{\rm vol}.\quad
\end{eqnarray}
They obey the following integrality
condition
\begin{equation}
  \label{eq:19}
  \frac{\sqrt2}{(2\pi\sqrt{\alpha'})^{p-1}} \int F_p = f_p\in\Z\ ,
  \qquad
  \frac{1}{(2\pi)^{2}\alpha'} \int H_3 = h_3\in\Z,
\end{equation}
so that the integer fluxes are related to the ones in (\ref{eq:18}) by
\begin{align}
  \label{eq:167}
  &f_0=-\sqrt2 2\pi\sqrt{\alpha'}m_0
  ,\qquad
  f_2^i=-\frac{\sqrt2 \k^{1/3}}{2\pi\sqrt{\alpha'}}m_i
  ,\qquad
  f_4^i=\frac{\sqrt2}{\k^{1/3}(2\pi\sqrt{\alpha'})^3}e_i
  ,\cr
  &\hspace{70pt} f_6=-\frac{\sqrt2}{(2\pi\sqrt{\alpha'})^5}e_0
  ,\qquad
  h_3=\frac{1}{(2\pi\sqrt{\alpha'})^2}p.
\end{align}

We will
split the field strengths
into
the background part and an excitations part.  They can then be written
as
\begin{eqnarray}
  \label{eq:20}
  H_3&=&H_3^{bg} + dB_2,\cr
  F_2&=&F_2^{bg} + dC_1 + m_0B_2,\cr
  F_4&=&F_4^{bg} + dC_3 - C_1\wedge dB_2 - \frac{m_0}{2} B_2\wedge B_2.
\end{eqnarray}
%
%
The background values of the fluxes are constrained by the tadpoles of
the different fields. These were analyzed in \cite{DeWolfe:2005uu},
where it was found that there is a unique tadpole for $C_7$ which
requires
\begin{equation}
  \label{eq:21}
  m_0 p =-\sqrt2\, 2\pi \sqrt{\alpha'}.
\end{equation}
In terms of the integer fluxes (\ref{eq:167}) this means $f_0h_3=2$, so
that there are four different possibilities,
$(f_0,h_3)=(1,2),(2,1),(-1,-2),(-2,-1)$.  All other fluxes are not
constrained by tadpoles.

The scalar potential was analyzed in detail in \cite{DeWolfe:2005uu},
and it was found
that by turning on such fluxes
$(e_0,e_i,m_0,m_i,p)$ the moduli are stabilized at values given by
\begin{eqnarray}
  \label{eq:22}
  \gamma_i&=&2(\kappa\sqrt3)^{1/3} \frac{1}{|\hat e_i|} \sqrt\frac{-5\hat e_1\hat e_2 \hat
    e_3}{3m_0\kappa},
  \cr
  b_i&=&\frac{m_i}{m_0},
  \cr
  e^{-\phi}&=&\frac43 \frac{1}{|p|} \(-\frac{12}{5}\frac{m_0\hat e_1\hat
    e_2\hat e_3}{\k}\)^{1/4},
  \cr
  \xi&=&\frac1p \(e_0+\frac{e_im_i}{m_0}+\frac{2\kappa m_1m_2m_3}{m_0^2}\),
\end{eqnarray}
with $\hat e_i \equiv e_i+\kappa m_jm_k/m_0$ (where $\{i,j,k\}=\{1,2,3\}$).
From the four dimensional point of view, this solution has a negative cosmological
constant
\begin{equation}
  \label{eq:23}
  \Lambda=-\frac{p^2}{2}\frac{\sqrt{3}}{\gamma_1\gamma_2\gamma_3},
\end{equation}
and we will consider the maximally symmetric solution of the resulting
four dimensional action, which is given by $AdS_4\times T^6/\Z_3^2$.

There are several things to note here regarding this solution. From
supersymmetry we get (see the next subsection) a constraint on the signs of the
fluxes
\begin{equation}
  \label{eq:169}
  {\rm sign}(m_0p)={\rm sign}(m_0e_i)=-,
\end{equation}
which also guarantees that the $\gamma_i$ and $e^{-\phi}$ are real. When we
take large values for the quantized fluxes, $f_4^i\gg 1$ (without making some
of them much larger than the others), we get to a regime with
large volume and weak coupling where we can trust our computation. Throughout
this paper we will work in this regime. We
also note that there is a non-singular solution with $e_0=m_i=0$,
which has no 2-form and 6-form background fluxes.

There are additional moduli localized
near the $C^3/\Z_3$ singularities. One can turn on $\hat F_2$ and
$\hat F_4$ fluxes on the corresponding localized cycles, which we denote, respectively, by $n_A$
and $f_A$ ($A=1,\cdots,9$ goes over the different singular
points). The blow up K\"ahler modes ${t_B}_A$ are then stabilized at
\begin{equation}
  \label{eq:24}
  {t_B}_A=\frac{n_A}{m_0} -i \sqrt{-\frac{10\hat f_A}{3\beta m_0}},
\end{equation}
where we defined $\hat f_A\equiv f_A+\beta n_A^2 / 2m_0$, and
the integer $\beta$ is the non-trivial triple intersection of the
twisted cycles. The values for $e^\phi$ and $\xi$ are modified
by these additional fluxes (the dilaton by a small amount when $f_4^i \gg 1$):
\begin{eqnarray}
  \label{eq:25}
  e^{-\phi}&=&\frac43 \frac1{|p|}
  \[
  \sqrt{-\frac{12}{5}\frac{m_0\hat e_1\hat e_2\hat e_3}{\k}}
  +
  \frac{3}{25}m_0^2\b\sum_A\(-\frac{10f_A}{3\b m_0}\)^{3/2}
  \]^{1/2}
  \cr
  \xi&=&\frac1p \(e_0+\frac{e_im_i+ \sum_A f_An_A}{m_0}+\frac{6\kappa m_1m_2m_3+\b\sum_An_A^3}{3m_0^2}\).
\end{eqnarray}

\subsection{Supersymmetry}
\label{sec:supersymmetry}

In this subsection we review how the background described above
satisfies the supersymmetry equations. We will write the background as
a warped product of a four-dimensional Anti de-Sitter space with
$T^6/\Z_3^2$, with the metric
\begin{equation}
  \label{eq:26}
  ds^2=e^{2A}h_{MN}dx^Mdx^N+g_{AB}dy^Ady^B,
\end{equation}
where $A=A(y)$ is the warp factor, $h_{MN}$ is the 4 dimensional AdS
metric and $g_{AB}$ is the metric on $T^6/\Z_3^2$.  We will use
the double spinor convention, which in type IIA amounts to writing the
Majorana Killing spinor as two Majorana Weyl spinors with opposite
chirality,
\begin{equation}
  \label{eq:27}
  \eps=\eps_++\eps_-, \quad
  \Gamma_{(10)} \eps_\pm=\pm\eps_\pm.
\end{equation}
We can decompose the ten dimensional Clifford algebra into the $4d\otimes6d$
algebras in the following way,
\begin{equation}
  \label{eq:28}
  \Gamma_{\ul\mu}=\gamma_{\ul\mu}\otimes\I,\quad
  \Gamma_{\ul m}=\gamma_{(4)}\otimes\hat\gamma_{\ul m},
\end{equation}
where the 4d gamma matrices are real and the 6d are purely imaginary
and antisymmetric. We denote by underlined indices the tangent space
flat indices. The Killing spinors also decompose as
\begin{eqnarray}
  \label{eq:29}
  \eps_+(x,y)&=& a\,\theta_+(x)\otimes\eta_+(y)+a^*\theta_-(x)\otimes\eta_-(y),\cr
  \eps_-(x,y)&=&b^*\theta_+(x)\otimes\eta_-(y)+b\,\theta_-(x)\otimes\eta_+(y),
\end{eqnarray}
where $\eta_+=\eta_-^*$ is the unique covariantly constant spinor on the
Calabi-Yau, while $\theta_+$, $\theta_-$ (with
$\bar\theta_+=\theta_-^TC$) are the Killing spinors on $AdS_4$
satisfying
\begin{equation}
  \label{eq:30}
  D_\mu\theta_+=\frac12\mu^*\gamma_\mu\theta_-,\quad
  D_\mu\theta_-=\frac12\mu\gamma_\mu\theta_+.
\end{equation}
The complex number $\mu$ is the value of the superpotential, so that
the cosmological constant of the $AdS_4$ space is given by
$\Lambda=-|\mu|^2$.

The spinor $\eta_+$ on the Calabi-Yau gives rise to an $SU(3)$
structure. Following
\cite{Grana:2004bg,Grana:2005sn,Martucci:2005ht,Koerber:2007jb} we can
write the two pure spinors as bispinors of $O(6,6)$ in the following
way
\begin{equation}
  \label{eq:31}
  \slash\Psi^+=a \eta_+\otimes b^*\eta_+^\dagger,\qquad
  \slash\Psi^-=a \eta_+\otimes b\eta_-^\dagger.
\end{equation}
Using the Clifford map, there is a one-to-one correspondence between
such bispinors and p-forms, given by
\begin{equation}
  \label{eq:32}
  C\equiv \sum \frac{1}{k!}
  C_{i_1,\ldots,i_k} dx^{i_1}\wedge\ldots\wedge dx^{i_k}
  \quad\longleftrightarrow\quad
  \slash C\equiv \sum \frac{1}{k!}
  C_{i_1,\ldots,i_k} \gamma_{\alpha\beta}^{i_1\ldots i_k}.
\end{equation}
Using this map, the pure spinors can also be represented by the almost
complex structure 2-form and the holomorphic 3-form,
\begin{equation}
  \label{eq:33}
  \Psi^+=\frac{a\bar b}{8}e^{-iJ},\qquad
  \Psi^-=-\frac{iab}{8}\Omega.
\end{equation}

Following these notations, the equations for preserved supersymmetry are
given by 
\cite{Grana:2005sn, Koerber:2007jb}
\begin{eqnarray}
  \label{eq:34}
  e^{-2A+\phi}(d+H\wedge)(e^{2A-\phi}\Psi_+)&=& 2\mu \Re{\Psi_-},\\
  e^{-2A+\phi}(d+H\wedge)(e^{2A-\phi}\Psi_-)&=& 3i\Im{\bar\mu\Psi_+}
  +dA\wedge\bar\Psi_-
  \cr &&+\frac{\sqrt2}{16}e^\phi\[(|a|^2-|b|^2)\hat F +i(|a|^2+|b|^2)
  \tilde F\],
  \cr&&
\end{eqnarray}
where $F=F_0+F_2+F_4+F_6$ are the modified RR fields defined as
\begin{equation}
  \label{eq:35}
  F= e^{-B} F^{\rm bg}+dC+H\wedge C,
\end{equation}
so that they obey the non-standard Bianchi identity $dF_n=-H\wedge
F_{n-2}$.

We solve these equations in Appendix \ref{sec:supersymm-equat-bulk},
finding that for supersymmetry to be preserved the Killing spinors
should have $b=-a^*$, and the moduli should obtain values as in (\ref{eq:22}).

\section{General Properties of the Dual Conformal Field Theory}
\label{sec:dual-cft}

In the previous section we described a solution of supergravity (and,
thus, of string theory) that includes a four dimensional AdS space.
According to the AdS/CFT correspondence
\cite{Maldacena:1997re,Witten:1998qj,Gubser:1998bc}, there is a
three dimensional conformal field theory which is the holographic dual
of this solution.  Many properties of this CFT can be calculated in a
simple manner from the supergravity solution. We will discuss these
properties in this section, including the central charge, dimensions
of operators and the global symmetries of the CFT. We will also
discuss D-branes wrapping cycles in the compact space to give
particles or strings on $AdS_4$.
Throughout this paper we will work only in the limit where all 4-form
fluxes are large, so that the string coupling is weak and the supergravity
approximation is good.

\subsection{The Central Charge}
\label{sec:central-charge}

We will begin by finding the central charge of the CFT from the
curvature of the AdS space.
There are various possible definitions of a central charge for three
dimensional CFTs, including the coefficient of the two-point function
of the stress-energy tensor, and the coefficient multiplying the
volume times the temperature squared in the entropy of the theory at finite temperature.
In the gravity approximation, all definitions give answers proportional to
$R_{AdS}^2/G_4$, where $G_4$ is the four dimensional Newton's constant, since
this is the coefficient (in units of $R_{AdS}$) of the four dimensional action,
so that all correlation functions are proportional to this.
Using our formulas from the previous section, we have
(up to constants)
\begin{equation}
  \label{eq:173}
  \frac{(R^{AdS}_4)^2}{G_4}
  =\frac{Vol(T^6/\Z_3^2)\,e^{-2\phi}\Lambda^{-1}}{\alpha'^4}
\propto \frac{(f_4^1 f_4^2 f_4^3)^{3/2}}{f_0^{5/2} h_3^4} \simeq
(f_4^1 f_4^2 f_4^3)^{3/2},
\end{equation}
since the 3-form and 0-form fluxes are numbers of order one.
In particular, if we take all the fluxes $f_4^i \sim N$, we find a central
charge scaling as $c \propto N^{9/2}$ (this was independently noted in \cite{Banks:2006hg}).

%
Equation (\ref{eq:173}) is reminiscent of the
formula for the central charge in the case of ${\cal N}=8$ $SU(N)$ SYM in 2+1
dimensions. In that case the central charge of the theory in the IR
(where it is dual to M theory on $AdS_4\times S^7$)
scales like $N^{3/2}$ (which is not understood in terms of any
effective field theory degrees of freedom). By analogy, this suggests
that in our case there may be some $N_{eff}=f_4^1 f_4^2 f_4^3$, namely
that if there is a UV description of
any sort it should include an order of $(f_4^1 f_4^2 f_4^3)^2$ degrees of
freedom. This is also suggested by the fact that these are the minimal
integer powers which are larger than those appearing in the central
charge (\ref{eq:173}). This UV description could be for instance an $SU(N_{eff})$
gauge theory, or an $SU(f_4^1)\times SU(f_4^2)\times SU(f_4^3)$ gauge theory with
matter in representations whose dimension is of order $N_{eff}^2$ (such
representations are consistent with asymptotic freedom in $2+1$ dimensions).

The analysis in \cite{Banks:2006hg} give some support to this suggestion.
It was argued there that after two T-dualities in the
directions of the first 2-torus, and in the limit of $f_4^1\rightarrow
\infty,\ f_4^{2,3}\ fixed$, the background should be lifted to M theory,
and resembles the near-horizon limit of $f_4^1$ M2-branes (at some
singularity).
In this case we see that the degrees of
freedom are renormalized from $O(1)*(f_4^1)^2$ in the theory on some
D2-branes (at the same singularity) to $O(1)*(f_4^1)^{3/2}$ in the theory on the M2-branes.

Below we will use another indicator for the number of branes in the
problem which will be the structure (and in particular the
dimensionality) of different branches of the moduli space. The moduli
space will be made out of holomorphic (in an appropriate sense)
D4-branes which wrap different 2-cycles of the torus. Our analysis of the
moduli space will be performed in the limit where all fluxes are
large, but since it preserves some supersymmetry it is natural to expect
that the same results for the form and dimension of the moduli space will
hold also in other limits (though we have not verified this directly). Assuming
this, we find (using our results derived
below) that for the scaling of \cite{Banks:2006hg} the
dimension of the largest branch of the moduli space will scale like $f_4^1$.
Indeed, this branch is described by the motion of
D4-branes wrapping the first $T^2$, which
become M2-branes (or D2-branes) after 2
T-dualities.

Our more general analysis below will show
that the dimension of the maximal branch of the moduli space scales like
$\max(f_4^i f_4^j),\ i\not=j$. The previous case is a special case of this.
Note that this might suggest that in a scaling limit in which two of
the fluxes (say, $f_4^1$ and $f_4^2$) become large while the third
remains finite, the theory resembles that of $N_{eff} \simeq f_4^1 f_4^2$
M2-branes. While the dimension of the moduli space and the
number of degrees of freedom are consistent with this suggestion, the
precise form of the moduli space is very different from what one would
obtain from any theory of $N_{eff}$ M2-branes.


\subsection{Global Symmetries}
\label{sec:global-symmetries}

As described above, the supergravity solution preserves a four dimensional
$\CN=1$ supersymmetry. By the AdS/CFT correspondence this maps to a three
dimensional $\CN=1$ superconformal symmetry, with two supersymmetry charges
and two superconformal charges.

In the AdS/CFT correspondence, the global symmetries of the CFT are related
to gauge symmetries of the gravitational theory. Such symmetries arise
from reductions of the supergravity fields on the compact space (or from
space-filling D-branes). The
simplest gauge fields are related to the ten dimensional metric, and are related to
the isometry group of the compactification manifold. In our case the
compact space is a Calabi-Yau manifold and thus has no isometry group.
So, we do not
get any gauge fields from the metric.
In addition to the metric, the RR 1-form and 3-form can also give rise to
gauge symmetries. In our background we have a non-trivial 0-form flux which
gives a mass to the 1-form (it is swallowed by the 2-form $B_2$ which
becomes massive). Thus, there is no gauge symmetry associated with the 1-form.
%
In order to get a 1-form gauge field from the 3-form we need to
integrate it over a 2-cycle. As the compactification manifold contains
three such untwisted 2-cycles, we obtain three commuting gauge fields. However,
since the 2-cycles are odd under the orientifolding, these gauge fields
are projected out by the orientifold. The gauge fields arising from the twisted
2-cycles are similarly projected out.

Thus, the conformal field theory that we are looking for does not have
any global symmetry (beyond the $\CN=1$ superconformal algebra, which does
not include any continuous R-symmetry group).

%

\subsection{Operators and Scalings}
\label{sec:operators-scalings}

Another basic property of a conformal field theory is the spectrum of
operators in the theory. The simplest operators are related to the
supergravity fields, and their dimensions are related to the masses so we can
easily find the spectrum. There are two mass scales for fields in the
supergravity. The first is the mass of the moduli, which can be computed
from their potential. This was written explicitly in \cite{DeWolfe:2005uu}
for some of the moduli, and it is easy to see that the others have the
same scaling.
In units of the four dimensional Planck scale $l_{p4}^2 \simeq G_4$ the moduli masses are
\begin{equation}
  \label{eq:58}
  m^2_{moduli}\sim(f_4^1 f_4^2 f_4^3)^{-3/2}l_{p4}^{-2}.
\end{equation}
The other mass scale in supergravity is the mass of the Kaluza-Klein modes, given by
the inverse radii of the compact tori,
\begin{equation}
  \label{eq:59}
  m^2_{KK}\sim \gamma_i^{-1} \sim (f_4^1 f_4^2 f_4^3)^{-3/2} f_4^i l_{p4}^{-2}.
\end{equation}
The dimensions of the corresponding operators are given using the AdS/CFT
correspondence as
\begin{equation}
  \label{eq:60}
  \Delta_{moduli}\sim m_{moduli}R_{AdS} \sim 1,
  \qquad
  \Delta_{KK}\sim m_{KK}R_{AdS} \sim \sqrt{f_4^i}.
\end{equation}
Thus, as in all other conformal field theories dual to theories with a
four dimensional supergravity approximation (implying a separation of
scales between the moduli and the KK modes), there is a small number
of operators with dimensions of order one, and all others have large
dimensions. The order one operators correspond to the eight moduli fields,
$\phi, \xi, b_i, v_i$.

\subsection{Wrapped Branes}
\label{sec:wrapped-branes}

Another type of operators in the field theory involves D$p$-branes
wrapped on $p$-cycles in the compact space, giving particles in the
$AdS_4$.
%
%
%
Since our background involves massive type IIA string theory, we cannot
have any D0-branes (which must have $f_0$ strings ending on them) or
D6-branes (which must have $f_0$ NS 5-branes ending on them); this is
related to the fact that the RR 1-form is swallowed by the NS-NS 2-form.
Naively we can have wrapped D2-branes or D4-branes on our 2-cycles or
4-cycles, but in fact the orientifold maps these to anti-D-branes, so it
is unlikely that any stable configurations of this type would exist.

We can also consider a $p$-brane wrapping a $(p-1)$-cycle, leading to
a string in $AdS_4$ (mapped to some type of flux tube in the conformal
field theory).
The only such possible configurations are a D4-brane wrapping a
3-cycle and an NS5-brane wrapped on a 4-cycle.
%
A D4-brane wrapped around the $\alpha_0$
cycle is mapped to an anti-brane by the orientifold, while a D4-brane wrapping the $\beta_0$
cycle is not a consistent configuration, since there is $H_3$-flux on
that 3-cycle, implying that such D4-branes must have D2-branes
ending on them. The same phenomenon arises for NS5-branes wrapped on the 4-cycles,
since these have 4-form flux.
Note that the fundamental string is also mapped to a string with opposite
orientation by the orientifold. Thus, we do not expect to have any stable
extended objects in our theory.

\section{Supersymmetric Domain Walls}
\label{sec:domain-walls}

In the next two sections we wish to study the moduli space of the conformal
field theory dual to the background described in section \ref{sec:supergravity-model}.
To describe the moduli space we need to find Lorentz-invariant configurations
with zero energy which have the same asymptotics as the solution described
above, but differ in the interior. Usually in the AdS/CFT correspondence
such configurations are described by supersymmetric branes sitting at some
value of the radial position, giving domain walls in AdS along which the
flux which the brane is charged under jumps.
Moving along the moduli space of these configurations
is described in the field theory side as giving non-trivial vacuum expectation
values to
operators. Such domain walls break half of the supersymmetry in the bulk;
in the conformal field theory they break the superconformal generators and
preserve the standard supersymmetry generators.


We will consider here D-brane domain walls,
given by D$p$-branes wrapping $(p-2)$-cycles in the compact space, and
sitting at fixed radial position in $AdS_4$.  For the configuration to
be supersymmetric (which is the same as having zero energy in the
field theory) these must obey some calibration condition
\cite{Koerber:2005qi}. We will find the supersymmetric cycles over
which D-branes can be wrapped by considering the $\kappa$-symmetry
equation. In Appendix \ref{sec:bps-condition} we will also verify
directly that these configurations are BPS states by considering the
DBI+CS action for the D-branes and checking that there is no force
acting on them. All of these equations are valid in the probe
approximation, in which the back-reaction of the D-brane on the
background is small. This approximation will be good in the limit of
large four-form fluxes that we are working in. Since in three
dimensional $\CN=1$ theories the moduli space is generally not
protected, we expect some potential along the moduli space to be
generated by corrections to our leading order approximation; however,
this potential is very small in the limit we are working in, so that
there will still be an approximate moduli space in the conformal field
theory.

The general supersymmetry condition for a D$p$-brane filling time plus
$q$ dimensions and wrapping a $(p-q)$-cycle in the compact directions
is the $\kappa$-symmetry equation \cite{Bergshoeff:1996tu}, which in
the double spinor notation can be written as in
\cite{Martucci:2005ht,Koerber:2007jb}:
\begin{equation}
  \label{eq:62}
  \hat \Gamma_{Dp} \eps_-=\eps_+,
\end{equation}
where
\begin{eqnarray}
  \label{eq:63}
  \hat \Gamma_{Dp} = \gamma_{\underline{0\ldots q}}
  \gamma_{(4)}^{p-q}\otimes \hat \gamma'_{(p-q)},
\end{eqnarray}
\begin{equation}
  \label{eq:64}
  \hat\gamma'_{(r)}=\frac1{\sqrt{{\rm det}(P[g]+\CF)}}
  \sum_{2l+s=r}
  \frac{\eps^{\alpha_1\ldots\alpha_{2l}\beta_1\ldots\beta_s}}{l!s!2^l}
  \CF_{\alpha_1\alpha_2}\ldots\CF_{\alpha_{2l-1}\alpha_{2l}}\hat\gamma_{\beta_1\ldots\beta_s}.
\end{equation}
Here, $P[\cdot]$ indicates the pullback of a bulk field onto the worldvolume of the
D-brane, and $\CF\equiv f+P[B]$ where $f$ is the field strength of the
gauge field on
the worldvolume of the D-brane, and we set $2\pi\alpha'=1$.

We can split the $\kappa$-symmetry equation into an equation in the
AdS space,
\begin{equation}
  \label{eq:65}
  \gamma_{\underline{0\ldots q}}\theta_+=\alpha^{-1}\theta_{(-)^{q+1}}
\end{equation}
for some constant $\alpha$, and an equation in the compact space
\begin{equation}
  \label{eq:66}
  b^{(*)^{p+1}}\hat\gamma'_{(p-q)}\eta_{(-)^{p+1}}=a^{(*)^{q+1}}\alpha\eta_{(-)^{q+1}},
\end{equation}
where $x^{(*)^n}$ is defined to be $x$ ($x^*$) for even (odd) values
of $n$ (and $a$ and $b$ were defined in (\ref{eq:29})). From these we can see (using the unitarity of the $\gamma$
matrices) that $\alpha$ must be a pure phase, and that the D-brane can
be supersymmetric only if $|a|=|b|$, which is indeed the case for our
background. For type IIA (even $p$) the internal equation can be brought
to the form
\begin{equation}
  \label{eq:67}
  b\hat\gamma'_{(p-q)}\eta_{+}=(-)^{p-q}a^{(*)^{p-q}}\alpha^*\eta_{(-)^{p-q}},
\end{equation}
from which one gets, as in \cite{Martucci:2005ht}, the following
calibration condition on the cycle which the D-brane wraps:
\begin{equation}
  \label{eq:68}
  \left.\left\{
      b^* P[e^{-iJ}]\wedge e^\CF
    \right\}\right|_{2k}
  =-a^*\alpha\sqrt{\det(P[g]+\CF)}\ d\sigma^1\wedge\ldots\wedge d\sigma^{2k}
\end{equation}
for D-branes wrapping even $2k$-cycles, and
\begin{equation}
  \label{eq:69}
  \left.\left\{
      b P[-i\Omega]\wedge e^\CF
    \right\}\right|_{2k+1}
  =a^*\alpha^*\sqrt{\det(P[g]+\CF)}\ d\sigma^1\wedge\ldots\wedge d\sigma^{2k+1}
\end{equation}
for D-branes wrapping odd $(2k+1)$-cycles. We denote the $n$-form part of
an expression by $\{\cdot\}|_n$.

We will next use this formalism to describe different configurations
of D-branes in this background and study their supersymmetry
properties. We begin by verifying that a D6-brane parallel to the
orientifold plane obeys the above equations. We then continue to
study the equation for D4-branes spanning domain walls in space-time.
After finding the general supersymmetric solution we will study the
special case of linear D-branes. In Appendix
\ref{sec:other-domain-walls} we show that there are no other types of
D-branes that lead to supersymmetric domain walls.

\subsection{A Space-Time Filling D6-Brane}
\label{sec:spacetime-filling-d6}

We start by considering a probe D6-brane filling the whole non-compact
$AdS_4$ space-time and wrapping a three-cycle in the compact space.
This is not a domain wall, but we use it to test our equations, since
we know that such a configuration carrying the same charges as the O6-plane
must be supersymmetric. The $AdS_4$ part of the
$\kappa$-symmetry equation (\ref{eq:65}) gives
\begin{equation}
  \label{eq:70}
  \alpha^{-1}\theta_+=\gamma_{\underline{0123}}\theta_+
  =i\gamma_{(4)}\theta_+=i\theta_+,
\end{equation}
so it fixes $\alpha=-i$.

Since the orientifold action is
\begin{equation}
  \label{eq:71}
  z_i\to-\bar z_i,
\end{equation}
the orientifold plane is located on $z_i=-\bar z_i$, and we wish to put
the D6-branes in the same position, so we can parameterize
the three compact coordinates of the D6-brane using the embedding
\begin{equation}
  \label{eq:72}
  \sigma_1=y_1,\quad\sigma_2=y_2,\quad\sigma_3=y_3.
\end{equation}
The induced metric on the worldvolume is
\begin{equation}
  \label{eq:73}
  ds^2=\sum_i \gamma_i (d\sigma^i)^2
\end{equation}
and the induced 3-form is
\begin{equation}
  \label{eq:74}
  P[\Omega]=\sqrt{\gamma_1\gamma_2\gamma_3}
  d\sigma_1\wedge d\sigma_2\wedge d\sigma_3.
\end{equation}

The right hand side of (\ref{eq:69}) is
\begin{equation}
  \label{eq:75}
  a^*\alpha^*\sqrt{\det(P[g]+\CF)}\ d\sigma_1\wedge d\sigma_2\wedge d\sigma_3
  =
  a^* i\sqrt{\gamma_1\gamma_2\gamma_3}\ d\sigma_1\wedge d\sigma_2\wedge d\sigma_3,
\end{equation}
and the left hand side of the equation is
\begin{equation}
  \label{eq:76}
  \left.\left\{
      b P[-i\Omega]\wedge e^\CF
    \right\}\right|_{3}
  =
  -ib \sqrt{\gamma_1\gamma_2\gamma_3}\
  d\sigma_1\wedge d\sigma_2\wedge d\sigma_3,
\end{equation}
so in order for the configuration to be supersymmetric we must have
$b=-a^*$, precisely as we found from the bulk supersymmetry in section \ref{sec:supersymmetry}.

\subsection{D4-Brane as a Supersymmetric Domain Wall}
\label{sec:d4-brane-as}

Next, consider a D4-brane extended as a domain wall in the AdS space and
wrapping a generic untwisted 2-cycle\footnote{One could also consider D4-branes wrapped around twisted 2-cycles, but it
seems that these are never supersymmetric in the presence of the 2-form
fluxes stabilizing the twisted sector moduli.}, in the cohomology class of $\sum N_iw_i$.
On such a domain wall, the fluxes jump by $f_4^i \to f_4^i\pm N_i$.  In
order to find the supersymmetric configuration we will solve the
$\kappa$-symmetry equation, starting as before with the $AdS_4$ part,
\begin{equation}
  \label{eq:77}
  \alpha^{-1}\theta_{-}=\gamma_{\underline{012}}\theta_+
  =\gamma_{\underline{012r}}\gamma_{\underline{r}}\theta_+
  =-\gamma_{\underline{r}}\gamma_{\underline{012r}}\theta_+
  =-i\gamma_{\underline{r}}\gamma_{(4)}\theta_+
  =-i\gamma_{\underline{r}}\theta_+.
\end{equation}
We choose the $AdS_4$ metric
\begin{equation}
  \label{eq:78}
  ds^2=\frac1{|\mu|}
  (dr^2+e^{2r}\eta_{\alpha\beta}dx^\alpha dx^\beta),
\end{equation}
where $\eta$ is a flat Minkowski metric. This is just the standard AdS
metric in the Poincar\'e patch, with the redefinition $r=-\ln(z)$. The
covariant derivative can be written as in \cite{Lu:1996rhb,Lu:1998nu},
\begin{equation}
  \label{eq:79}
  D_\alpha=\pd_\alpha
  +\frac12 |\mu| e^r\gamma_{\ul\alpha}\gamma_{\ul r}.
\end{equation}
We are interested in the Poincar\'e supercharges, obeying $\pd_\alpha
\theta_\pm=0$, so using (\ref{eq:30}) we get
\begin{equation}
  \label{eq:80}
  \frac12 |\mu| e^r\gamma_{\ul\alpha}\gamma_{\ul r}\theta_+
  =\frac12\mu^*\gamma_\alpha\theta_-
  =\frac12\mu^*e^r\gamma_{\ul\alpha}\theta_-
\end{equation}
\begin{equation}
  \label{eq:81}
  \gamma_{\ul r}\theta_+
  =\frac{\mu^*}{|\mu|}\theta_-
  =-{\rm sign}(p)\frac{b}{\bar b}\theta_-.
\end{equation}
Plugging this into (\ref{eq:77}) we find $\alpha=-{\rm sign}(p)i\frac{\bar b}{b}$.

To solve the internal part of the $\kappa$-symmetry equation we need
to choose how to wrap the D4-brane. We start with the simplest case
where the D4-brane wraps the torus $z_1$. We can choose the embedding
\begin{equation}
  \label{eq:82}
  \sigma^1=x^1,\quad \sigma^2=y^1,
\end{equation}
with the induced metric being
\begin{equation}
  \label{eq:83}
  \gamma_1 (d\sigma^1)^2 +\gamma_1 (d\sigma^2)^2,
\end{equation}
and the pullback of $J$ given by
\begin{eqnarray}
  \label{eq:84}
  P[J]=\gamma_1 d\sigma^1\wedge d\sigma^2.
\end{eqnarray}

Plugging into the supersymmetry condition (\ref{eq:68}) we have on
the right-hand side
\begin{equation}
  \label{eq:85}
  -a^* \alpha \sqrt{\det(P[g]+\CF)}d\sigma^1\wedge d\sigma^2
  =i {\rm sign}(p)\frac{a^* b^*}{b} \gamma_1 d\sigma^1\wedge d\sigma^2
  =-{\rm sign}(p)i b^* \gamma_1 d\sigma^1\wedge d\sigma^2,
\end{equation}
while on the other side we have
\begin{equation}
  \label{eq:86}
  \left.\left\{
      b^* P[e^{-iJ}]\wedge e^\CF
    \right\}\right|_{2}
  =-ib^*P[J]
  =-ib^* \gamma_1 d\sigma^1\wedge d\sigma^2.
\end{equation}
We see that when the background value of $p$ is positive the
configuration is supersymmetric.\footnote{Recall that the signs of $p$
  and $e_i$ are the same (\ref{eq:57}).} When $p$ is negative one can
take the same embedding and flip its orientation such that
$\sigma^1=y^1$ and $\sigma^2=x^1$, to get a supersymmetric
configuration. This is just an anti-D4-brane instead of a D4-brane. We
see that depending on the sign of the background fluxes, the
supersymmetric brane is either a D4-brane or an anti-D4-brane.

Note that in the above we assumed $\CF=P[B]+f=0$. As the
$\kappa$-symmetry equation only depends on $\CF$, the result won't
change if we have a non-trivial background $F_2$ (which generates also
a non-trivial background $B$)
as long as we turn on fluxes on the worldvolume $f=-P[B]$. If the
worldvolume flux is different than this value there will be an
additional contribution to both sides of the equation. In the right
hand side $\CF$ appear only inside the square root so it will change
only the absolute value while keeping the phase unchanged. In
contrast, the left hand side is proportional to $\CF-iJ$ and so will
change its phase. We thus conclude that the configuration is
supersymmetric only for $\CF=0$.

A different type of cycle the D4-branes can wrap is a twisted
cycle at a fixed point. When we go away from the singular limit by
turning on 2-form flux on these cycles, the background fluxes and
values of the moduli change, see equation (\ref{eq:25}). However the
$\kappa$-symmetry equations are only sensitive to changes in the bulk
supercharges, that is to the relation between $a$ and $b$ which
remains unchanged. Thus, by turning on the appropriate worldvolume
flux on D4 branes wrapping the twisted cycles such that $\CF=0$ as
before we get additional supersymmetric configurations.
We will not consider these configurations in detail, since their
contribution to the dimension of the moduli space is finite in
the large flux limit.

\subsection{Generic D4-Brane Configuration}
\label{sec:generic-d4}

Since the linear embedding described in the previous subsection cannot
be realized for generic values of the $N_i$, we
will now analyze the most general supersymmetric case of a D4-brane wrapping a
generic (untwisted) surface.
We will use the complex coordinates $z_a=x_a+iy_a$ in space-time as in
(\ref{eq:1}) and define the worldvolume complex coordinate to be
$\s=\s_1+i\s_2$ with the same complex structure. The position of the
D4-brane can be written as
\begin{equation}
  \label{eq:101}
  z^a=z^a(\s,\bar\s).
\end{equation}
The induced metric is given by
\begin{eqnarray}
  \label{eq:102}
  g_{\s\s}&=&\sum_{a=1,2,3} \gamma_a \pd z_a \pd\bar z_a,
  \cr
  g_{\s\bar\s}&=&g_{\bar\s\s}=\sum_{a=1,2,3}
  \frac12\gamma_a (\pd z_a\bar\pd \bar z_a+\bar\pd z_a \pd\bar z_a),
  \cr
  g_{\bar\s\bar\s}&=&\sum_{a=1,2,3} \gamma_a \bar\pd z_a \bar \pd\bar z_a,
\end{eqnarray}
so the right-hand side of the $\k$ equation is proportional to
\begin{eqnarray}
  \label{eq:103}
  \sqrt{\(\sum_a\frac12\gamma_a (\pd z_a\bar\pd \bar z_a+\bar\pd z_a \pd\bar
    z_a)\)^2
  -\sum_a \gamma_a \pd z_a \pd\bar z_a
  \sum_b \gamma_b \bar\pd z_b \bar\pd\bar z_b}.
\end{eqnarray}
This should be equal to the pullback of the almost complex structure,
which gives
\begin{eqnarray}
  \label{eq:104}
  \sum_a\frac12\gamma_a (\pd z_a\bar\pd \bar z_a-\bar\pd z_a \pd\bar z_a).
\end{eqnarray}
Taking the squares of both sides and equating we get
\begin{eqnarray}
  \label{eq:105}
  0&=&
  \frac12\sum_{ab} \g_i\g_j
  |\pd z_a \bar\pd z_b-\pd z_b \bar\pd z_a|^2,
\end{eqnarray}
which vanishes if and only if
\begin{equation}
  \label{eq:106}
  \pd z_a \bar\pd z_b=\pd z_b \bar\pd z_a.
\end{equation}

In order to understand the meaning of this result, let's consider $z_1$
and $z_2$. We start by defining a new variable $\w=z_1(\s,\bar\s)$. We
have
\begin{eqnarray}
  \label{eq:107}
  d\w&=&\pd z_1 d\s+\bar\pd z_1d\bar\s, \cr
  d\bar\w&=&\pd\bar z_1 d\s+\bar\pd\bar z_1d\bar\s,
\end{eqnarray}
and
\begin{eqnarray}
  \label{eq:108}
  d\s&=&\frac{\bar\pd\bar z_1 d\w-\bar\pd z_1d\bar\w }
  {\bar\pd\bar z_1\pd z_1-\bar\pd z_1\pd\bar z_1},\cr
  d\bar\s&=&\frac{\pd z_1 d\w+\pd\bar z_1d\bar\w}
  {\bar\pd\bar z_1\pd z_1-\bar\pd z_1\pd\bar z_1}.
\end{eqnarray}
We now can write
\begin{equation}
  \label{eq:109}
  \frac{\pd z_2}{\pd \bar w}
  =\frac{\pd \s}{\pd\bar\w}\pd z_2+\frac{\pd\bar\s}{\pd\bar\w}\bar\pd z_2
  =\frac{1}{\bar\pd\bar z_1\pd z_1-\bar\pd z_1\pd\bar z_1}
  (-\pd z_2 \bar\pd z_1+\pd z_1 \bar\pd z_2)
\end{equation}
which vanishes according to (\ref{eq:106}). We see that the
supersymmetry condition can be understood as the statement that the
three coordinates $z_a$ can be written as holomorphic functions of
each other. In other words, supersymmetry is equivalent to the
requirement that the worldvolume wraps a cycle that can be written as
the zero locus of two holomorphic functions of the coordinates.

\subsubsection{Linear D4-Brane}
\label{sec:linear-d4}

We will study now a simple class of configurations, in which the
embedding of the D-brane can be chosen to be a linear map. We can write the embedding
as
\begin{equation}
  \label{eq:87}
  x^i=a^i \sigma^1+b^i \sigma^2+\alpha^i,\qquad y^i=c^i\sigma^1+d^i\sigma^2+\beta^i.
\end{equation}
Two of the six parameters $\alpha^i,\ \beta^i$ can be absorbed into a
shift in $\sigma_1,\ \sigma_2$, while the others parameterize the
moduli of the position of the D4-brane. We also need to check that
this embedding keeps the periodicity of the tori. The identifications
on $\sigma_1,\sigma_2$ are
\begin{equation}
  \label{eq:94}
  (\sigma_1,\sigma_2) \simeq (\sigma_1+1,\sigma_2)
  \simeq (\sigma_1+\frac12,\sigma_2+\frac{\sqrt3}{2})
\end{equation}
and similarly for the $(x_i,y_i)$ pairs. Under the first
transformation, we get
\begin{equation}
  \label{eq:95}
  (x_i,y_i)\to (x_i+a_i,y_i+c_i).
\end{equation}
For these two points to be identified we must have $c_i=\frac{\sqrt3}{2}
m_i$ and $a_i=\frac {m_i} 2 +n_i$ for some integers $m_i,n_i$.
The second transformation acts as
\begin{equation}
  \label{eq:96}
  (x_i,y_i)\to
  (x_i+\frac{a_i}{2}+\frac{\sqrt3}{2}b_i
  ,y_i+\frac{c_i}{2}+\frac{\sqrt3}{2}d_i),
\end{equation}
which gives us the restrictions $d_i=\tilde m_i-\frac {m_i} 2$ and
$b_i=\frac{1}{\sqrt3} (2\tilde n_i+\tilde m_i-\frac {m_i} 2-n_i)$ (with
integers ${\tilde m}_i, {\tilde n}_i$). We
are now able to express $a,b,c,d$ in terms of four integers
$m,n,\tilde m,\tilde n$.
The wrapping numbers $N_i$ are given by
\begin{equation}
  \label{eq:88}
  N_i=\frac{\int_{\sigma^1,\sigma^2}dx^idy^i}{\int_{x^i,y^i}dx^idy^i}
  =\det\(\[
  \begin{array}{c c}
    a^i&b^i\\
    c^i&d^i\\
  \end{array}
  \]\)
  \frac{\int_{\sigma^1,\sigma^2}d\sigma^1d\sigma^2}
  {\int_{x^i,y^i}dx^idy^i}
  =a^id^i-b^ic^i=n^i\tilde m^i - \tilde n^i m^i.
\end{equation}

Plugging the embedding into the supersymmetry equations (\ref{eq:106})
we get
\begin{eqnarray}
  \label{eq:97}
  &&m_j\tilde m_i-m_i\tilde m_j+n_i\tilde n_j-n_j\tilde n_i=0,\\
  &&m_i\tilde m_j-m_j\tilde m_i+n_i\tilde m_j-m_j\tilde n_i-n_j\tilde m_i+m_i\tilde n_j=0,\\
  &&n_i\tilde m_i-m_i\tilde n_i = N_i.
\end{eqnarray}
We can solve the first two equations for $\tilde n,\ \tilde m$, and
plugging into the third we get
\begin{equation}
  \label{eq:98}
  r_{ij}\equiv\frac{N_i}{N_j}=\frac{m_i^2+m_in_i+n_i^2}{m_j^2+m_jn_j+n_j^2}.
\end{equation}

It turns out that not all charges $N_i$ may be realized by a single
linear D4-brane of the type described above, since there is not always
an integer solution to (\ref{eq:98}). To see this, we will now
prove some things about this ratio. First, note that
\begin{equation}
  \label{eq:99}
  4(m^2+mn+n^2)=3(n+m)^2+(n-m)^2\equiv 3x^2+y^2
\end{equation}
so we can write the equation as
\begin{equation}
  \label{eq:100}
  r_{ij}=\frac{3x_i^2+y_i^2}{3x_j^2+y_j^2}
\end{equation}
with integer $x_i,y_i$.
Next, we will show that a number that can be written as $N=3x^2+y^2$ has
an even power for the factor of $2$ in its prime decomposition. Then, the
ratio of two such numbers must also have an even power
for the $2$ in its prime decomposition (if the ratio is a fraction its
prime decomposition is the one coming from the prime decompositions of the
numerator and denominator).

We will prove this by induction, showing that if $N$ is divisible by
$2^{2n+1}$ then it is also divisible by $2^{2n+2}$. For $n=0$, if both
$x,y$ are even, $N$ is obviously divisible by $4$. Else for $N$ to be
even both $x,y$ have to be odd, i.e. of the form $x=2a+1,\ y=2b+1$. We
then get $N=3x^2+y^2=3(4a^2+4a+1)+4b^2+4b+1=4(3a^2+b^2+3a+b)+4$ which
is divisible by 4.

We next consider general $n$. Again, since $N$ is even, $x,y$ are both
even or both odd. In the first case we can divide the entire equation
by 4 and reduce it to the case with $n-1$. For the latter case, we can
write again $N=3x^2+y^2=3(4a^2+4a+1)+4b^2+4b+1=4(3a(a+1)+b(b+1)+1)$,
which after division by 4 is an odd number, specifically it is not a
multiple of $2^{2n+1}$, so this case cannot arise.

\section{The Geometry of the Moduli Space}
\label{sec:geom-moduli-space}

We have seen that the background of section
\ref{sec:supergravity-model} allows for supersymmetric domain walls,
described by D4-branes wrapped on 2-cycles. Over each domain wall the
4-form fluxes jump according to the number of times the domain wall is
wrapped over each cycle. When we go far away from the domain walls, we
arrive at a background with specific values for the 4 form fluxes.
However there are many different configurations of domain walls which
result in the same background in the interior of AdS space (beyond all
the domain walls).  For example, we can take one D-brane wrapped $N_i$
times over the $i$'th cycle, or several branes whose total wrapping
number is $N_i$. From this we see that the moduli space may be
composed of many different branches. The parameterization of each
branch includes the radial position of the domain walls, so each
branch is a cone, and all the branches are connected at the origin
(when all the domain walls go to the horizon of AdS space). Naively,
the full moduli space is made out of all configurations of D4-branes
carrying total wrapping numbers equal to the total fluxes $f_4^i$
(some of the D4-branes can of course sit at the origin). However, it
is not completely clear that this is true, since our approximations
break down when the 4-form fluxes $f_4^i$ become small (and it is
certainly not clear if there is an $AdS_4$ solution when one of the
fluxes vanishes). Nevertheless, we expect that this naive approach
will be a useful tool for counting the dimension of the moduli space
at large $f_4^i$.

Note that often configurations made out of different sets of
D4-branes (with the same total wrapping number) can be connected
without the need to send some of the branes to the origin of the
moduli space. When two D4-branes intersect, new light degrees of
freedom arise at their intersection point which may deform the
configuration and smooth it into a configuration of a single D4-brane
with the same total flux.

We will begin by considering a simple branch of the moduli space where
there is only one D4-brane wrapping a simple cycle. We will study it
in detail and describe its global structure. We will then go on to
describe some properties of the general moduli space.  Specifically,
we will parameterize the different branches, and estimate the
dimension of a generic branch.

\subsection{The Moduli Space of a Single D4-Brane}
\label{sec:low-energy-action}

We start with the simplest branch of the moduli space, which includes
branes with wrapping numbers $(N_1,N_2,N_3)=(1,0,0)$. For these values
we can have only one possible configuration of domain walls, which
consists of a single D4-brane wrapping the first $T^2$ inside the
compact space. The geometry of its moduli space can be simply read
from its low-energy effective action. We consider the D4-brane to be
located at specific values of $r,u^2,v^2,u^3,v^3$ and embedded as
\begin{equation}
  \label{eq:118}
  t=\xi^0,\quad
  x^1=\xi^1,\quad
  x^2=\xi^2,\quad
  v^1=\xi^3,\quad
  u^1=\xi^4.\quad
\end{equation}

We begin by assuming that the D4-brane is away from all fixed points of the
orbifold and orientifold. The DBI action is given (up to quadratic
order in the fields) by
\begin{align}
  \label{eq:119}
  \CL_{DBI}=&-\mu_4\int d^5\xi \exp{-\phi}\sqrt{-g_{ik}}
  \cr
  \approx&-\mu_4\int d^5\xi \exp{-\phi}\frac{r^3}{R^3}\gamma_1\[
  1+\frac12 G_{\i\k}\partial_iX^{\i} \partial_kX^\k g^{ik}
  +\frac14\CF_{ik}\CF_{i'k'}g^{ii'}g^{kk'}
  \]
\end{align}
where $g_{ik}=\frac{\partial X^I}{\partial \xi^i}\frac{\partial
  X^K}{\partial \xi^k}G_{IK}$ is the induced metric on the D-brane,
and $G_{IK}$ is the ten dimensional metric which we now write in the form (with $R=R_{AdS}$)
\begin{equation}
  \label{eq:185}
  ds^2=\frac{r^2}{R^2}\(dt^2+(dx^1)^2+(dx^2)^2\)
  +\frac{R^2}{r^2}dr^2+\sum_{i=1}^{3}\gamma_i\((du^i)^2+(dv^i)^2\).
\end{equation}
We use $i,k$ to denote the worldvolume indices, $\i,\k$ are transverse
coordinates and $I,K$ denote full ten dimensional indices. Reducing
the action on the torus we take the fields $r,u^2,v^2,u^3,v^3$ to
depend only on $t,x_1,x_2$. The 5 dimensional gauge field $A$ can be
expanded as \footnote{For $F_2\neq0$ we need to take the gauge field
  to have non vanishing background flux so that
  $\CF=0$. This doesn't change the rest of the analysis.}
\begin{equation}
  \label{eq:120}
  A=\hat A+a_1du^1+a_2dv^1,
\end{equation}
giving rise to two
Wilson lines $a_i$, and a three dimensional gauge field, $\hat A$,
which can be dualized to another scalar $*_3d\hat A=d\phi$. Using $\int
du^1dv^1=\frac{\sqrt{3}}{2}$ we get the three dimensional action
\begin{align}
  \label{eq:121}
  \CL_{DBI}=&
  -\mu_4\frac{\sqrt3}{2}\int d^3\xi
  \exp{-\phi}\frac{r}{R}\gamma_1\bigg[
  \frac{r^2}{R^2}+
  \frac12 \partial_i\phi\partial^i\phi
  +\frac12 \frac{1}{\gamma_1}\partial_ia_1\partial^ia_1
  +\frac12 \frac{1}{\gamma_1}\partial_ia_2\partial^ia_2
  \cr&\qquad
  +\frac12 \frac{R^2}{r^2}\partial_i r\partial^i r
  +\frac12 \gamma_2 \partial_iu^2\partial^iu^2
  +\frac12 \gamma_2 \partial_iv^2\partial^iv^2
  +\frac12 \gamma_3 \partial_iu^3\partial^iu^3
  \cr&\qquad
  +\frac12 \gamma_3 \partial_iv^3\partial^iv^3
  \bigg],
\end{align}
with indices raised and lowered using the flat metric.

The Chern-Simons term is
\begin{equation}
  \label{eq:122}
  \sqrt2\mu_4\int \CC_5,
\end{equation}
where $\CC_5=C_5+C_3\wedge \CF_2+\frac12C_1\wedge \CF_2\wedge \CF_2
+\frac16m_0\omega_5$, with $d\omega_5=\CF_2\wedge \CF_2\wedge \CF_2$.
This can be written as an integral of a 6-form, $\CF_6=d\CC_5$, over
the volume bounded by the D4-brane. In our background only $F_6$
contributes, and using the calculation in Appendix
\ref{sec:bps-condition}, we have
\begin{equation}
  \label{eq:123}
  \sqrt2\mu_4\int F_6=\mu_4\int dtdx_1dx_2 e^{-\phi}\frac{r^3}{R^3}\gamma_1\[
  du^1\wedge dv^1
  +\frac{\gamma_2}{\gamma_1}du^2\wedge dv^2
  +\frac{\gamma_3}{\gamma_1}du^3\wedge dv^3\],
\end{equation}
which becomes
\begin{equation}
  \label{eq:124}
  \sqrt2\mu_4\int F_6=\mu_4\int d^5\xi e^{-\phi}\frac{r^3}{R^3}\[ \g_1
  +\gamma_2
  \(\frac{\pd u^2}{\pd\xi^3}\frac{\pd v^2}{\pd\xi^4}
  -\frac{\pd u^2}{\pd\xi^4}\frac{\pd v^2}{\pd\xi^3}\)
  +\gamma_3
  \(\frac{\pd u^3}{\pd\xi^3}\frac{\pd v^3}{\pd\xi^4}
  -\frac{\pd u^3}{\pd\xi^4}\frac{\pd v^3}{\pd\xi^3}\)\]
\end{equation}
when we use our specific embedding of the D4-brane. When we
compactify, we assume that no fields depend on the compact coordinates, so the only
term that contributes to the low-energy effective action is the
constant, which is canceled with the constant term in the DBI part.

We can also redefine the radial coordinate to be
$\rho=\sqrt{2\sqrt3\frac{\mu_4}{g_s}\g_1R}\sqrt{r}$ so that the action is
given by
\begin{align}
    \CL_{DBI}=&
  -\int d^3\xi \bigg[ \frac{1}{2} \partial_i \rho \partial^i \rho
  +\frac12\frac{\rho^2}{4R^2}\big(
  \partial_i\phi\partial^i\phi
  +\gamma_1^{-1}\partial_ia_1\partial^ia_1
  +\gamma_1^{-1}\partial_ia_2\partial^ia_2
  \cr&\qquad
  +\gamma_2 \partial_iu^2\partial^iu^2
  +\gamma_2 \partial_iv^2\partial^iv^2
  +\gamma_3 \partial_iu^3\partial^iu^3
  +\gamma_3 \partial_iv^3\partial^iv^3
  \big) \bigg].
\end{align}
This describes an 8 dimensional moduli space which is a cone (with
radial coordinate $\rho$) over a 7 dimensional space parameterized
by $\phi,a_1,a_2,u^2,v^2,u^3,v^3$.

To study the global structure of the moduli space we will consider now
each of the scalar fields. Starting with the dual scalar, $\phi$, one
can see that it is actually periodic. In a 3d YM theory, whose action
is given by
\begin{equation}
  \label{eq:125}
  \int d^3x \sqrt{g}\frac{1}{4g_{YM}^2}f_{\mu\nu}f^{\mu\nu}
  =\frac{1}{g_{YM}^2}\int f\wedge *f,
\end{equation}
the electric charge inside an $S^1$ is given by
\begin{equation}
  \label{eq:126}
  Q_e=\frac{1}{g_{YM}^2} \int_{S^1} *f
  =\frac{1}{g_{YM}^2} \int_{S^1} d\phi
  =\frac{1}{g_{YM}^2} (\phi(2\pi)-\phi(0)).
\end{equation}
Since the field values $\phi(0)$ and $\phi(2\pi)$ are the same, and
$Q_e$ are integers we have
\begin{equation}
  \label{eq:127}
  \phi\simeq \phi+g_{YM}^2.
\end{equation}
In our case $g_{YM}^2=\frac{g_s}{\mu_4}\frac{2}{\sqrt3 \gamma_1}$.

The Wilson lines are also periodic fields. Performing a gauge
transformation $A\to A+d\Lambda$ with $\Lambda=c_1u^1+c_2v^1$, on a
torus of complex structure $\tau$, shifts the Wilson lines by $a_i\to
a_i+c_i$.
Since $e^{i\Lambda}$ must be periodic under the identifications of the
coordinates given by $(u^1,v^1) \sim (u^1+1,v^1) \sim
(u^1+\Re\tau,v^1+\Im{\tau})$, we need
\begin{equation}
  \label{eq:129}
  c_1=2\pi n_1,\qquad
  c_1\Re\tau+c_2\Im{\tau}=2\pi n_2,
\end{equation}
for integers $n_1$ and $n_2$. These are solved for integral
linear combinations of
\begin{align}
  \label{eq:130}
  &\{c_1=2\pi,\qquad c_2=2\pi\frac{1-\Re\tau}{\Im\tau}\}
  \cr
  &\{c_1=0,\qquad\ \  c_2=2\pi\frac{1}{\Im\tau}\}.
\end{align}
Under the corresponding gauge transformation the fields do not
change so we must identify these points on the moduli space of the
Wilson lines. Therefore we get (using $\tau=e^{i\pi/3}$)
\begin{align}
  (a_1,a_2) \sim (a_1+2\pi,a_2+\frac{2\pi}{\sqrt3}) \sim (a_1,a_2+\frac{4\pi}{\sqrt3}).
\end{align}
This is a torus with complex structure $\tau=e^{i\pi/3}$ in the
coordinate $\tilde z=\frac{\sqrt3}{4\pi}(a_2+ia_1)$.

Thus, we see that the moduli space has the structure of a cone with a seven
dimensional base $S^1\times(T^2)^3$ (before imposing the orbifold
and orientifold identifications), where the circumference of the $S^1$
is
\begin{eqnarray}
  \label{eq:131}
  2\pi R_\phi
  =\frac{2}{\sqrt3}\frac{g_s}{\mu_4}\frac{1}{\gamma_1},
\end{eqnarray}
and the complex structure of the three tori are all $\tau=e^{i\pi/3}$,
while their volumes are
\begin{equation}
  \label{eq:132}
  \(\frac{4\pi}{\sqrt3}\)^2\frac1{\gamma_1}
  ,\qquad
  \gamma_2
  ,\qquad
  \gamma_3.
\end{equation}


The above analysis was for a D4-brane located at a generic point,
where it is separated from its images.  However, we can consider also
a D4-brane located at the fixed points of the $T^6/\Z_3^2$. We start
with the non fractional brane (and obtain the fractional ones from it
via higgsing). The D-brane wraps $u^1,v^1$, and it can sit at a fixed
point on the other two tori.  There are three such points, distinct
after all identifications. In the covering space of the orbifold
action, $T^6$, the D4-brane has nine copies, which are divided into
three separate groups of three coincident branes. To study the moduli
space we need to consider the transformation of the Chan-Paton
indices. Under the first $\Z_3$ The fields transform as
\begin{eqnarray}
  \label{eq:133}
  \phi_{ij}  &\to&  \alpha^{2(i-j)} \phi_{ij},
  \cr
  r_{ij}  &\to&  \alpha^{2(i-j)} r_{ij},
  \cr
  a_{ij}=(a_1+ia_2)_{ij}   &\to&   \alpha^{2(1+i-j)} a_{ij},
  \cr
  z^2_{ij}=(u^2+iv^2)_{ij} &\to& \alpha^{2(1+i-j)} z^2_{ij},
  \cr
  z^3_{ij}=(u^3+iv^3)_{ij} &\to& \alpha^{2(1+i-j)} z^3_{ij},
\end{eqnarray}
where $\alpha=e^{i\pi/3}$. The invariant fields are then
\begin{eqnarray}
  \label{eq:134}
  \phi=
  \begin{pmatrix}
    \phi_{00}&0&0\\
    0&\phi_{11}&0\\
    0&0&\phi_{22}
  \end{pmatrix},
  \cr
  r=
  \begin{pmatrix}
    r_{00}&0&0\\
    0&r_{11}&0\\
    0&0&r_{22}
  \end{pmatrix},
  \cr
  a=
  \begin{pmatrix}
    0&a_{01}&0\\
    0&0&a_{12}\\
    a_{20}&0&0
  \end{pmatrix},
  \cr
  z^i=
  \begin{pmatrix}
    0&z^i_{01}&0\\
    0&0&z^i_{12}\\
    z^i_{20}&0&0
  \end{pmatrix}.
\end{eqnarray}

The moduli space is determined by considering commuting matrices,
since the scalar potential contains terms with commutators. We then
find two branches. On one branch the fields $a,z^2,z^3$ vanish, and
$\phi,r$ can have any value, giving rise to 3 scalars each. This
describes the D4-brane and its images at the fixed point as
fractional branes with the corresponding gauge group of $U(1)^3$, each
at a different radial position. The second is when $a,z^2,z^3$ are
generic and $\phi$ and $r$ are proportional to the identity matrix, in
which case the D-brane is away from the fixed points and has some
non-trivial Wilson line. Here the gauge group is broken back to a single
$U(1)$.  The position and Wilson lines are given by
$\sqrt[3]{z_{01}^iz_{12}^iz_{20}^i}$ and
$\sqrt[3]{a_{01}a_{12}a_{20}}$, respectively. This just spans locally a
$\Z_3$ singularity. The global identifications are just as in the case
away from the fixed points.

We also need to consider the effect of the orientifold action,
$\Omega(-1)^{F_L} \s$, where $\s$ is the spacetime involution $z_i\to
-\bar z_i$, $F_L$ is the worldsheet left moving fermion number and
$\Omega$ is the worldsheet parity reversal.  The action on our fields
is
\begin{eqnarray}
  \label{eq:136}
  \phi_{i,j}        &\to& -\phi_{-j,-i}
  \cr
  r_{i,j}           &\to& r_{-j,-i}
  \cr
  a_{i,j}           &\to& \bar a_{-j,-i}
  \cr
  z^2_{i,j}         &\to& -\bar z^2_{-j,-i}
  \cr
  z^3_{i,j}         &\to& -\bar z^3_{-j,-i}.
\end{eqnarray}
We then get the following degrees of freedom:
\begin{eqnarray}
  \label{eq:137}
  \phi_{22}=-\phi_{11},\qquad \phi_{00}=0,\cr
  r_{22}=r_{11},\qquad r_{00},\cr
  a_{01}=\bar a_{20},\qquad a_{12}=\bar a_{12},\cr
  z^i_{01}=-\bar z^i_{20},\qquad z^i_{12}=-\bar z^i_{12}.
\end{eqnarray}
The D-brane position is now $i|\sqrt[3]{z^i_{01}z^i_{12}z^i_{20}}|$ so
it can move only along the O-plane. To move out of this plane the
D-brane must meet its image and so we need a pair of such D-branes.
Similarly, the Wilson line is $|\sqrt[3]{a_{01}a_{12}a_{20}}|$.

\subsection{Generic Properties of the Moduli Space}
\label{sec:gener-prop-moduli}

In the previous section we found that supersymmetric domain wall
configurations are described by a holomorphic curve. Here we will
provide a more detailed description of a generic branch of this
type, and explain how to count its dimension (in the limit of large
charges). The main tools that we will use are the Bezout and
Bernstein theorems, which we will review, which will be used to
calculate the wrapping numbers of a generic branch. We will be
interested primarily in the branch of largest dimension, and
examine how this maximal dimension scales with the wrapping numbers.

\subsubsection{Mathematical Preliminaries}
\label{sec:math-prel}

We will now introduce some mathematical theorems that will help us
count the number of solutions for a system of generic polynomial
equations. More details can be found in \cite{cox}. The basic
theorem that answers this question is Bezout's Theorem:
\vspace{10pt}

\begin{tabular}[c]{@{\hspace{0pt}}p{390pt}}
 If the equations $f_1=\ldots=f_n=0$ have degree $d_1,\ldots,d_n$ and
 finitely many solutions in $\mathbb{CP}^n$, then the number of solutions
 (counted with multiplicity) is $d_1\cdots d_n$.
\end{tabular}
\vspace{10pt}

\noindent
This theorem holds for any polynomials $f_i$ in the complex projective
space.

We will be interested, however, in polynomials in $\mathbb{C}^n$.
Given such polynomials $f_i\in\mathbb{C}[x_1,\ldots,x_n]$ with terms
of total degree up to $d_i$, we can always add an additional variable,
$z$, making all terms of total degree $d_i$. We can then view them as
equations in $\mathbb{CP}^n$,
and then we can apply Bezout's theorem and find the
number of solutions.  We will assume generic polynomials, so that one
can assume no solutions at $z=0$. By gauging $z=1$ we can reduce
each solution in the projective space to a solution in $\mathbb{C}^n$.
We thus have this version of Bezout's theorem
\vspace{10pt}

\begin{tabular}[c]{@{\hspace{0pt}}p{390pt}}
  Given $n$ generic polynomials $f_1,\ldots,f_n$, if the equations
  $f_1=\ldots=f_n=0$ have maximal total degree $d_1,\ldots,d_n$ and
  finitely many solutions in $\mathbb{C}^n$, then the number of solutions
  (counted with multiplicity) is $d_1\cdots d_n$.
\end{tabular}
\vspace{10pt}

\noindent
Here we assume that the polynomials are generic in the sense that all terms
with degree up to $d_i$ appear with a non vanishing coefficient in the
polynomial $f_i$.

In our case we will have polynomials that are generic in a different
sense than what was used in the previous case. The polynomial $f_i$
will contain all terms that are up to order $d_i^a$ in each variable
$x_a$.\footnote{Actually, one can relax this condition, but this will
  not change the scaling behavior of the dimensionality of the moduli
  space as we will discuss in the next subsection.}  This is obviously
less generic than needed for Bezout's theorem so we will need to use
Bernstein's theorem, a generalization of Bezout's theorem. We will
start by introducing some concepts used in Bernstein's theorem.

\begin{figure}[tb]
  \centering
  \includegraphics[scale=0.5]{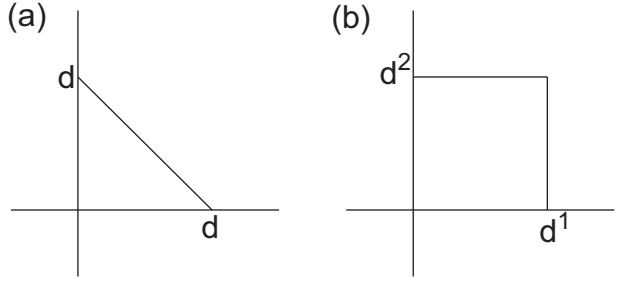}
  \caption{The Newton polytopes of (a) a polynomial with highest total
    degree $d$. (b) a polynomial with each $x_a$ having highest degree
    $d_a$.}
  \label{fig:polytopes}
\end{figure}

Let $f\in \mathbb{C}[x_1,\ldots,x_n]$ be a polynomial in $n$ variables. We can
describe it by a set of points in the positive integer lattice
$\Z^n_{\ge0}$, each point corresponding to a monomial.
We can write
\begin{equation}
  \label{eq:175}
  f=\sum_{\alpha\in\Z^n_{\ge0}} c_\alpha x^\alpha,
\end{equation}
and the set of points is given by
\begin{equation}
  \label{eq:176}
  \CA=\{\alpha\in\Z^n_{\ge0} : c_\alpha\neq0\}.
\end{equation}
This set of points can be used to define the Newton polytope of $f$,
given by the convex hull of $\CA$
\begin{equation}
  \label{eq:177}
  {\rm NP}(f)={\rm Conv}(\CA)=\left\{\sum_{\alpha\in\CA}\lambda_\alpha \alpha
  : \lambda_\alpha \ge0,\sum_{\alpha\in\CA}\lambda_\alpha =1\right\}.
\end{equation}
A polynomial is said to be generic if $c_\alpha\neq0$ for any
lattice point $\alpha$ inside its Newton polytope.  As an example, for
$n=2$, the Newton polytope for a polynomial with all terms of order up
to $d$ is given by the triangle in figure \ref{fig:polytopes}(a). The
Newton polytope of a polynomial with terms up to $d^a$ in the variable
$x_a$ is given by the square in figure \ref{fig:polytopes}(b).

There are two operations that can be carried out on polytopes in
$\R^n$ in order to generate new ones. Let $P,Q$ be polytopes in $\R^n$
and let $\lambda\ge0$ be a real number.
\begin{enumerate}
\item The Minkowski sum of $P$ and $Q$ denoted $P+Q$, is
\begin{equation} 
P+Q=\{p+q : p\in P \ {\rm and}\  q\in Q\},
\end{equation}
where $p+q$ denotes the usual vector sum in $\R^n$
\item The polytope $\lambda P$ is defined by
\begin{equation}
\lambda P=\{\lambda p : p\in P\},
\end{equation}
where $\lambda p$ is the usual scalar multiplication on $\R^n$.
\end{enumerate}

We will also define the mixed volume of a collection of polytopes
$P_1,\cdots,P_n$, denoted
\begin{equation}
MV_n(P_1,\cdots,P_n)
\end{equation}
to be the coefficient of the monomial
$\lambda_1\lambda_2\cdots\lambda_n$ in the volume of the polytope
$P=\lambda_1P_1+\cdots+\lambda_nP_n$.

Using the notions introduced above, we can now write Bernstein's
theorem as follows \cite{cox,ber}
\vspace{10pt}

\begin{tabular}[c]{@{\hspace{0pt}}p{390pt}}
  Given polynomials $f_1,\ldots,f_n$ over $\C$ with finitely
  many common zeroes in $(\mathbb{C}^*)^n$, let $P_i=NP(f_i)$ be the Newton
  polytope of $f_i$ in $\R^n$. Then the number of common zeros of the
  $f_i$ in $(\mathbb{C}^*)^n$ is bounded above by the mixed volume
  $MV_n(P_1,\ldots,P_n)$. Moreover, for generic choices of the
  coefficients in the $f_i$, the number of common solutions is exactly
  $MV_n(P_1,\ldots,P_n)$.
\end{tabular}
\vspace{10pt}

\noindent
For the two cases in $\R^2$ described in figure \ref{fig:polytopes} it
is simple to calculate the mixed volume. Polynomials $f_1,f_2$ with
all terms up to order $d_1,d_2$ have triangular Newton polytopes, as
in figure \ref{fig:polytopes}(a), and their mixed volume is given by
\begin{equation}
MV_n(P_1,P_2)=d_1 d_2,
\end{equation}
while polynomials with terms up to
order $d_i^1$ in $x_1$ and $d_i^2$ in $x_2$ have square Newton
polytopes as in figure \ref{fig:polytopes}(b), for which
\begin{equation}
MV_n(P_1,P_2)=d_1^1d_2^2+d_1^2d_2^1.
\end{equation}

\subsubsection{The Branches of Moduli Space}
\label{sec:setting-up-problem}

Next we will use Bernstein's theorem to calculate the properties of
the moduli space for a generic D4-brane (or several D4-branes)
wrapping a 2-cycle on the compact space. We have seen that the
supersymmetry condition requires the embedding of the D4-brane to be
holomorphic, so the 2-cycle is given by a set of two holomorphic
equations in the $z_i$.  Since the $z_i$ are doubly periodic the
holomorphic equations should be periodic as well. The most general
elliptic function over a torus with complex structure $\tau$ can be
written in terms of the periodic Weierstrass functions $w_i\equiv
\wp(z_i|\tau)$ and their derivatives $w_i'\equiv \wp'(z_i|\tau)$.
For the purposes of Berenstein's theorem we will treat these
variables as independent and add to the set of polynomials $f_i$ the
relations
\begin{equation}
  \label{eq:138}
  w_i'^2-(4w_i^3+g_2(\tau) w_i+g_3(\tau))=0,\ \ \ i=1,2,3.
\end{equation}
A general supersymmetric D-brane will thus be located at the zeros of
(\ref{eq:138}) and of two holomorphic polynomials of the form
\begin{equation}
  \label{eq:139}
  P(w_i,w_i')=Q(w_i,w_i')=0.
\end{equation}
We can restrict the polynomials to have terms only up to first order
in $w_i'$, since higher powers can be removed using the relations
(\ref{eq:138}).  We will take the highest degree of the variable
$w_i$ in $P$ and $Q$ to be $p_i,q_i$, respectively\footnote{The
D-brane configuration is described by the vanishing locus of a set
of polynomials where the highest degree of each parameter is
constrained separately. Perhaps one can relax this condition
by considering zeros of this set of equations that enter from
infinity or from zeros of $w_i$. On the torus side, in the
computations below, we can always avoid such points.}.

Given a set of such polynomials, which describe a D4-brane, we will
use Bernstein's theorem, applied to subsets of this set, to count
the wrapping number of the D-brane on the different cycles. Consider
for example $N_1$ -- the number of times the brane wraps the
$z_1$ cycle. We will evaluate $N_1$ by fixing a value of $z_1$ and
then counting how many solutions there are to the equations for this
$z_1$ (for a generic $z_1$). Fixing $z_1$ means that we fix both
$w_1$ and $w_1'$ which satisfy the constraint (\ref{eq:138}) for
$i=1$. This leaves us with 4 polynomials in the variables
$w_2,w_3,{w'}_2,{w'}_3$, on which we apply Bernstein's theorem. The
number of solutions to these equations is
\begin{equation}
  \label{eq:140}
  N_1\sim p_2q_3+p_3q_2,
\end{equation}
and similarly for permutations of $\{1,2,3\}$.

Recall that we fix $N_{1,2,3}$ and count the dimensionality of the
moduli space for this set of $N$'s. We are interested in finding the
values of $p_i$ and $q_i$ for which we obtain the largest
dimensionality. The dimension of the moduli space for a given set of
$p_i$ and $q_i$ can be estimated by the number of different
monomials in the two polynomials, which is $8p_1p_2p_3+8q_1q_2q_3$.
However, the actual dimension of the moduli space is smaller, since
different pairs of polynomials might have the same zero locus. If we
assume $q_i<p_i$, then any multiple of $Q$ with degree smaller than
$p_i$ can be removed from $P$. We are thus left with a moduli space
of dimension
\begin{equation}
  \label{eq:141}
  D\sim8(p_1p_2p_3+q_1q_2q_3-(p_1-q_1)(p_2-q_2)(p_3-q_3)).
\end{equation}

To summarise, we have found that the moduli space describing a domain
wall across which the flux jump by $(N_1,N_2,N_3)$ units of flux,
consists of different branches each parametrized by a set of 2
polynomials with degrees satisfying (\ref{eq:140}). The dimension of
such a branch is given by (\ref{eq:141}).

\subsubsection{The Maximal Moduli Space}
\label{sec:maximal-moduli-space}

It is interesting from the point of view of the dual field theory to
understand how the dimension of the moduli space scales as we take
large wrapping numbers, $N_i>>1$ (which should still be much smaller
than the fluxes since we are using the probe approximation). For this
we will find the dimension of the maximal branch with given wrapping
numbers. We can use (\ref{eq:140}) to solve for $q_i$ in terms of the
$p_i$ for given values of the fluxes,
\begin{equation}
  \label{eq:160}
  q_1=\frac{-N_1p_1+N_2p_2+N_3p_3}{2p_2p_3}
\end{equation}
and similarly for $q_2,q_3$. Since the $q_i$'s are positive, this give some
non-trivial condition on the $p_i$'s. The requirement $q_i<p_i$ then
leads to the inequalities
\begin{eqnarray}
  \label{eq:161}
  -N_1p_1+N_2p_2+N_3p_3 < 2p_1p_2p_3,\cr
  N_1p_1-N_2p_2+N_3p_3 < 2p_1p_2p_3,\cr
  N_1p_1+N_2p_2-N_3p_3 < 2p_1p_2p_3,\cr
\end{eqnarray}
which can be brought to the form
\begin{equation}
  \label{eq:162}
  N_1<2p_2p_3
\end{equation}
and its permutations. In the same way we can get $N_1>2q_2q_3$
and its permutations.

We can now use our results in the equation (\ref{eq:141}) for $D$. The
term with three $p_i$'s cancels. For the terms of the form $ppq$ we
can use (\ref{eq:160}) to see that they scale like $N_ip_i < N_iN_j$,
since $p_i$ can be just as large as $N_j$ ($j\neq i$). Next, we have
$pqq<pN$ so these terms are also smaller than $N_iN_j$. Finally, the
term with three $q_i$'s is $qqq<Nq$ and scales as the other terms.
Terms with less than three $p$'s or $q$'s are smaller for the same
reasons. We thus conclude that for large fluxes the dimension of
moduli space behaves as
\begin{equation}
  \label{eq:163}
  D \leq \sum_{i\neq j}N_iN_j.
\end{equation}
We can actually find a configuration which saturates this
bound on the dimensionality of moduli space. For instance, if all $N_i$
are of the same order, then by choosing all $q_i\sim
1$ we get $p_i\sim\sum N_j$, in which case we get $D\sim \sum_{i\neq
  j}N_iN_j$.

The previous analysis was done under the assumption that each $q_i$ is
smaller than $p_i$ so that we can eliminate terms in the polynomial
$P$ using $Q$ thus reducing the dimension of moduli space. However it
is possible that this is not the case. If one of the $q_i$'s is larger we
need to take all monomials, and the dimension of the moduli space is
$D\sim p_1p_2p_3+q_1q_2q_3$. We will assume that $q_1$ is the large
$q$ so that $q_1>p_1, q_{2,3}<p_{2,3}$. We find
\begin{eqnarray}
  \label{eq:164}
  q_1>p_1 &\to& -N_1p_1+N_2p_2+N_3p_3>2p_1p_2p_3,\cr
  q_2<p_2 &\to& N_1q_1-N_2q_2+N_3q_3>2q_1q_2q_3.
\end{eqnarray}
From the first inequality we get that $p_1p_2p_3< N_iN_j$ and
from the second one we get that also $q_1q_2q_3< N_iN_j$ so that we
arrive again to the same conclusion (\ref{eq:163}) as before.

In addition to the directions in the moduli space that change the two
polynomials and control the embedding of the D-brane, there are additional
dimensions of the moduli space related to Wilson lines. The number of
Wilson lines is related to the 1 dimensional homology group of the
Riemann surface the D-brane is wrapping. We can try to estimate this
number as follows. We can think of the polynomials $P(w_i,w_i'),Q(w_i,w_i')$ as
polynomials in a projective space by adding a new variable $\lambda$
and making all terms have the same weight of $p=\sum p_i$ and $q=\sum
q_i$. The relations (\ref{eq:138}) are then of weight $3$. It is then
possible using algebraic geometry methods to calculate the Euler
characteristic of the complete intersection of these 5 polynomials to
be
\begin{equation}
  \label{eq:165}
  \chi=-27pq(2-p-q).
\end{equation}
As before, we have $ppq\sim N_ip_j$ and $qqp\sim N_iq_j$ so that we
have $\chi\sim N_iN_j$. Since the number of Wilson lines is just the
genus, it scales as the Euler characteristic, and we get that
\begin{equation}
  \label{eq:166}
  D_{Wilson}\sim \sum_{i\neq j}N_iN_j,
\end{equation}
as before. We thus conclude that the total dimension of the moduli space
with given wrapping numbers scales in the same fashion,
\begin{equation}
  \label{eq:186}
  D_{total}\sim \sum_{i\neq j}N_iN_j.
\end{equation}

We note that this behavior may point us towards an
$SU(f_4^1)\times SU(f_4^2)\times SU(f_4^3)$ gauge theory, as the dimension of the
moduli space can than be viewed as coming from the degrees of freedom of strings
sitting in the bifundamental representation of any two $SU(N)$ factors.


\section*{Acknowledgements}

We would like to thank K. Blum, G. Engelhard, Y. Hochberg, S. Kachru,
A. Lawrence, J. Louis, R. Minasian, A. Sever, E.  Silverstein, J.
Simon, A.  Tomasiello and T.  Volanski for useful discussions.  This
work was supported in part by the Israel-U.S.  Binational Science
Foundation, by a center of excellence supported by the Israel Science
Foundation (grant number 1468/06), by a grant (DIP H52) of the German
Israel Project Cooperation, by Minerva, by the European network
MRTN-CT-2004-512194, and by a grant from G.I.F., the German-Israeli
Foundation for Scientific Research and Development.

\appendix

\section{Supersymmetry Equations in the Bulk}
\label{sec:supersymm-equat-bulk}

In this appendix we solve the equations for supersymmetry in the bulk
for the background discussed in section \ref{sec:supergravity-model}.
We find the unbroken spinors and the values for the stabilized moduli.

The equations for preserved supersymmetry are given by
\cite{Grana:2005sn, Koerber:2007jb}
\begin{eqnarray}
  \label{eq:34}
  e^{-2A+\phi}(d+H\wedge)(e^{2A-\phi}\Psi_+)&=& 2\mu \Re{\Psi_-},\\
  e^{-2A+\phi}(d+H\wedge)(e^{2A-\phi}\Psi_-)&=& 3i\Im{\bar\mu\Psi_+}
  +dA\wedge\bar\Psi_-
  \cr &&+\frac{\sqrt2}{16}e^\phi\[(|a|^2-|b|^2)\hat F +i(|a|^2+|b|^2)
  \tilde F\],
  \cr&&
\end{eqnarray}
where $F=F_0+F_2+F_4+F_6$ are the modified RR fields defined as
\begin{equation}
  \label{eq:35}
  F= e^{-B} F^{\rm bg}+dC+H\wedge C,
\end{equation}
so that they obey the non-standard Bianchi identity $dF_n=-H\wedge
F_{n-2}$.

Plugging our background into the first equation we get
\begin{eqnarray}
  \label{eq:36}
  &&H\wedge \Psi_+ =  2\mu \Re{\Psi_-}
  \cr  &&
  \Rightarrow -p\beta_0\wedge \frac{a\bar b}{8} e^{-iJ}
  =  2\mu \Re{-\frac{iab}{8}\Omega}
  \cr &&
  \Rightarrow -pa\bar b \beta_0 =  2\mu \Im{ab\Omega}
  =\sqrt2\mu\frac{\sqrt{\gamma_1\gamma_2\gamma_3}}{3^{1/4}} (\Re{ab}
  \beta_0+\Im{ab} \alpha_0),
  \cr&&
\end{eqnarray}
where
$\frac{\sqrt{\gamma_1\gamma_2\gamma_3}}{3^{1/4}\sqrt2}\beta_0=\Im{\Omega}$
and
$\frac{\sqrt{\gamma_1\gamma_2\gamma_3}}{3^{1/4}\sqrt2}\alpha_0=\Re{\Omega}$.
The solution is $\Im{ab}=0$ and
\begin{eqnarray}
  \label{eq:37}
  -pa\bar b &=& \sqrt2\mu \frac{\sqrt{\gamma_1\gamma_2\gamma_3}}{3^{1/4}} ab
  \Rightarrow\cr
  \mu&=&-\frac{p}{\sqrt2}\frac{3^{1/4}}{\sqrt{\gamma_1\gamma_2\gamma_3}}
  \frac{\bar b}{b}
  \Rightarrow
  \Lambda=-|\mu|^2=\frac{p^2}{2}\frac{\sqrt{3}}{\gamma_1\gamma_2\gamma_3}.
\end{eqnarray}

In the second equation we use for the left hand side
\begin{equation}
  \label{eq:39}
  H\wedge\Psi_-=-p\beta_0\wedge\frac{-iab}{8}
  \frac{\sqrt{\gamma_1\gamma_2\gamma_3}}{3^{1/4}}
  \frac{1}{\sqrt2}(\alpha_0+i\beta_0)
  =-\frac{ipab}{8\sqrt2} \frac{\sqrt{\gamma_1\gamma_2\gamma_3}}{3^{1/4}}
  \alpha_0\wedge\beta_0.
\end{equation}
This equation can be split according to the rank of the forms that
appear in it. The zero-form part of the equation is
\begin{equation}
  \label{eq:40}
  0=3i\Im{\frac{\bar\mu a\bar b}{8}}+\frac{\sqrt2}{16}e^\phi
  \[(|a|^2-|b|^2)\hat F_0-i(|a|^2+|b|^2)*_6\hat F_6\].
\end{equation}
The first term is proportional to $\Im{ab}$ so it vanishes. The real part implies
\begin{equation}
  \label{eq:41}
  |a|=|b|,
\end{equation}
assuming a non-vanishing value for $\hat F_0$,
while the imaginary part of this equation requires $\hat F_6=0$
which gives, using (\ref{eq:35}),
\begin{eqnarray}
  \label{eq:42}
  0&=&\int \hat F_6
  =\int H\wedge C_3 +\hat F_6^{\rm bg}-\hat F_4^{\rm bg}\wedge B
  +\frac12\hat F_2^{\rm bg}\wedge B\wedge B
  -\frac16F_0^{\rm bg}\wedge B\wedge B\wedge B
  \cr
  &=&p\xi - e_0 - e_ib_i - \frac12 \kappa b_i b_j m_k
  + \kappa m_0 b_1 b_2 b_3,
\end{eqnarray}
with $\{i,j,k\}$ being summed over all permutations of $\{1,2,3\}$.

The two-form part is
\begin{equation}
  \label{eq:43}
  0=3i\Im{\frac{\bar\mu a\bar b}{8}(-iJ)}
  +\frac{\sqrt2}{16}e^\phi\[(|a|^2-|b|^2)\hat F_2+i(|a|^2+|b|^2)\tilde F_2\],
\end{equation}
with
\begin{eqnarray}
  \label{eq:44}
  &&J=\frac{\gamma_i}{2}(\kappa\sqrt3)^{-1/3} \omega_i,
  \cr&&
  \hat F_2 = -m_i w_i+m_0b_iw_i,
  \cr&&
  \tilde F_2=*_6 \hat F_4=-\hat e_i *\tilde\omega^i
  =\frac{2 \hat e_i\gamma_i^2}{\gamma_1\gamma_2\gamma_3}
  \(\frac{\sqrt3}{\kappa^2}\)^{1/3}\omega_i,
\end{eqnarray}
where we used
\begin{eqnarray}
  \label{eq:45}
  \hat F_4&=&\int H\wedge C_1 +\hat F_4^{\rm bg}-\hat F_2^{\rm bg}\wedge B
  +\frac12\hat F_0^{\rm bg}\wedge B\wedge B
  \cr
  &=&(e_i+\k (m_jb_k+m_kb_j) -\k m_0b_jb_k)\tilde w^i
  =\hat e_i \tilde w^i.
\end{eqnarray}

Again, (\ref{eq:43}) splits into real and imaginary parts. The
real part vanishes since $|a|=|b|$,
while the imaginary part reduces to
\begin{eqnarray}
  \label{eq:48}
  0&=&3i \frac{-\bar\mu a\bar b}{8}(J)
  +i\frac{\sqrt2}{16}e^\phi(|a|^2+|b|^2)\tilde F_2
  \cr
  &=&i\frac38 \frac{\gamma_i}{2}(\kappa\sqrt3)^{-1/3} \omega_i
  \frac{p}{\sqrt2}\frac{3^{1/4}}{\sqrt{\gamma_1\gamma_2\gamma_3}}
  ab
  +i e^\phi |a|^2
  \frac{2\sqrt2\hat e_i\gamma_i^2}{8 \gamma_1\gamma_2\gamma_3}
  \(\frac{\sqrt3}{\kappa^2}\)^{1/3}\omega_i
\end{eqnarray}
which gives
\begin{eqnarray}
  \label{eq:49}
  \frac{e^{-2\phi}\hat e_i\gamma_i}{\sqrt{\prod_i e^{-2\phi}\hat e_i\gamma_i}}
  =-\frac{3^{11/12}}{8}p\k^{1/3}\frac{ab}{|a|^2}
\end{eqnarray}
(with no summation over $i$).
This can be solved to give
\begin{equation}
  \label{eq:50}
  e^{-2\phi}\g_i=\frac{64}{3^{11/6}}\frac{\hat e_1\hat e_2\hat
  e_3}{\hat e_ip^2\k^{2/3}},
\end{equation}
where we used $\Im{ab}=0$ and $|a|=|b|$ to get $b=\pm a^*$ or
$ab=\pm|a|^2$.
We will later see that we must take the minus sign for the background
to be supersymmetric.

The 4-form part of the equation is
\begin{equation}
  \label{eq:51}
  0=3i\Im{\frac{\bar\mu a\bar b}{8}\frac12(-iJ)^2}
  +\frac{\sqrt2}{16}e^\phi\[(|a|^2-|b|^2)\hat F_4+i(|a|^2+|b|^2)\tilde F_4\].
\end{equation}
Just as before, the first term vanishes, and since we have
$\tilde F_4=*\hat F_2$ we get
\begin{equation}
  \label{eq:52}
  0=\int_{[w_i]} F_2=-m_i+m_0 b_i \ \Rightarrow\ b_i=\frac{m_i}{m_0}.
\end{equation}
Plugging this back into (\ref{eq:45}) we get $\hat
e_i=e_i+\k\frac{m_im_j}{m_0}$.

Finally, the 6-form part is
\begin{equation}
  \label{eq:53}
  H\wedge\Psi_-=3i\Im{\frac{\bar\mu a\bar b}{8}\frac16(-iJ)^3}
  +\frac{\sqrt2}{16}e^\phi\[(|a|^2-|b|^2)\hat F_6+i(|a|^2+|b|^2)\tilde F_6\].
\end{equation}
We use (\ref{eq:39}) and
\begin{eqnarray}
  \label{eq:54}
  &&
  \tilde F_6 = *_6 \hat F_0 = -m_0 *_61
  \cr&&
  \frac16J^3=\frac{i}{8}\Omega\wedge\bar\Omega
  =\frac18\frac{\gamma_1\gamma_2\gamma_3}{\sqrt3}\alpha_0\wedge\beta_0
  =*1
\end{eqnarray}
to get
\begin{eqnarray}
  \label{eq:55}
  0&=&-H\wedge\Psi_-+3i\Im{\frac{\bar\mu a\bar b}{8}\frac16(-iJ)^3}
  +\frac{\sqrt2}{16}e^\phi\[(|a|^2-|b|^2)\hat F_6+i(|a|^2+|b|^2)\tilde F_6\]
  \cr&=&
  \frac{ipab}{8\sqrt2} \frac{\sqrt{\gamma_1\gamma_2\gamma_3}}{3^{1/4}}
  \alpha_0\wedge\beta_0
  +3i\frac{\bar\mu a\bar b}{8}\frac18\frac{\gamma_1\gamma_2\gamma_3}{\sqrt3}
  \alpha_0\wedge\beta_0
  -\frac1{16}e^\phi i2|a|^2\sqrt2m_0\frac18\frac{\gamma_1\gamma_2\gamma_3}{\sqrt3}
  \alpha_0\wedge\beta_0
  \cr&=&
  \frac{ipab}{8\sqrt2} \frac{\sqrt{\gamma_1\gamma_2\gamma_3}}{3^{1/4}}
  \alpha_0\wedge\beta_0
  -3i
  \frac{p}{\sqrt2}\frac{3^{1/4}}{\sqrt{\gamma_1\gamma_2\gamma_3}}
  \frac{ab}{8}\frac18\frac{\gamma_1\gamma_2\gamma_3}{\sqrt3}
  \alpha_0\wedge\beta_0
  \cr&&
  -\frac1{16}e^\phi i2|a|^2\sqrt2m_0\frac18\frac{\gamma_1\gamma_2\gamma_3}{\sqrt3}
  \alpha_0\wedge\beta_0
  \cr&=&
  \frac58\frac{ipab}{8\sqrt2} \frac{\sqrt{\gamma_1\gamma_2\gamma_3}}{3^{1/4}}
  \alpha_0\wedge\beta_0
  -\frac1{16}e^\phi i2|a|^2\sqrt2m_0\frac18\frac{\gamma_1\gamma_2\gamma_3}{\sqrt3}
  \alpha_0\wedge\beta_0,
\end{eqnarray}
which gives us
\begin{equation}
  \label{eq:56}
  e^{4\phi}\sqrt{\prod_i e^{-2\phi}\g_i}=\frac{5\cdot3^{1/4}}{2}\frac{p}{m_0}\frac{ab}{|a|^2}.
\end{equation}
Using (\ref{eq:50}) we can solve for $e^\phi$ and $\g_i$. We
know that the O6-plane generates a tadpole that is canceled by the
fluxes $m_0$ and $p$ according to (\ref{eq:21}), so that we find ${\rm
  sign}(m_0p)=-$.  We thus must have, for supersymmetry to hold,
$b=-a^*$ as stated in the main text. The results we got for the moduli
agree with (\ref{eq:22}).

We note that equations (\ref{eq:49}), (\ref{eq:56}) give us the following
conditions on the signs of the background fluxes,
\begin{eqnarray}
  \label{eq:57}
  {\rm sign}(pe_i)=+,
  \quad
  {\rm sign}(pm_0)=-.
\end{eqnarray}

\section{BPS Condition}
\label{sec:bps-condition}

In this appendix we will consider the supersymmetric domain wall
solutions found in section \ref{sec:domain-walls}. Such a
supersymmetric configuration should be a BPS state and therefore feel
no radial force. We will verify this fact directly by considering the
D-brane effective action. (This was independently verified in
\cite{Koerber:2007jb}.) In the supersymmetric configuration the
gravitational force coming from the DBI term will be canceled against
the RR force coming from the WZ term, related to the charge of the
D-brane.

The D-branes extend along a $2+1$ dimensional surface in $AdS_4$
parallel to the boundary at constant $r$, and wrap a $(p-2)$-cycle in
the compact space.
Their world-sheet action in the string frame is given by
\begin{equation}
  \label{eq:143}
  I_{brane}=I_{DBI}+I_{WZ}~,
\end{equation}
where
\begin{equation}
  \label{eq:144}
  I_{DBI}=-\mu_p\int d^pxe^{-\phi}\sqrt{-\det\(G+\CF\)}
\end{equation}
is a Dirac-Born-Infeld type action in the string frame, $\mu_p$ is the
D-brane tension and
\begin{equation}
  \label{eq:145}
  \CF=f+P[B]~.
\end{equation}
$I_{WZ}$ is the following Wess-Zumino (WZ) type action
\begin{equation}
  \label{eq:146}
  I_{WZ}=\sqrt{2}\mu_p\int\(\CC\wedge
  e^\CF+m_0\omega\)~,
\end{equation}
where
\begin{equation}
  \label{eq:147}
  \CC=\sum_{i=0}^9\CC_{i}~,\qquad d\omega=e^{\CF}
\end{equation}
and $m_0$ is the massive type IIA mass parameter. The $\sqrt{2}$ is
due to the different normalization of the RR fields we use (following
\cite{DeWolfe:2005uu}).

As in \cite{SW}, the brane action has two contributions which depend
on the radial location of the brane in $AdS_4$. One contribution is
proportional to the brane area $A$ and comes from the DBI action, the
other is proportional to the volume enclosed by the brane $V$ and
comes from the WZ action.  Next, we are going to evaluate the different
terms for wrapped D4-branes. 
We will assume $\CF=0$.


The $D4$-brane domain walls wrap a two-cycle in
the compact space. In our background there is a 3-dimensional basis
for the untwisted 2-cycles given by $[\omega_i],\ i=1,2,3.$ For
simplicity we consider wrapping the two-cycle $[\omega_1]$ in
$T^6/Z_3^2$.

\vspace{10pt}
\noindent {\bf DBI}
\label{sec:dbi}
\vspace{5pt}

In this case we have
\begin{equation}
  \label{eq:148}
  \sqrt{-\det\ G}=\gamma_1 du^1dv^1 \frac{r^3}{R_{\rm AdS}^3}dtdx^1dx^2,
\end{equation}
so we have
\begin{equation}
  \label{eq:149}
  I_{\rm DBI}=\mu_4\int d^5xe^{-\phi}\sqrt{-\det\ G}
  =\mu_4a\gamma_1 e^{-\phi}{r^3 \over R_{\rm AdS}^3}\int d^3x
  =\mu_43^{-{13\over 12}}2^\frac72 5^\frac14
  {\kappa^\frac13 E^\frac34\over e_1pm_0^\frac14}
  \ {r^3 \over R_{\rm AdS}^3}a\int d^3x~,
\end{equation}
where we define
\begin{equation}a\equiv\int_{[\omega_1]}du^1 dv^1~.\end{equation}

\vspace{10pt}
\noindent {\bf WZ}
\label{sec:wz}
\vspace{5pt}

The only non-zero contribution to the WZ term is given by
\begin{equation}
  \label{eq:150}
  \mu_4\sqrt{2}\int_{W_5}\CC_5
  =\mu_4\sqrt{2}\int_{{\rm Vol}(W_5)} F_6,
\end{equation}
where $W_5$ is the D4-brane worldvolume wrapping a 2-cycle in the
compact space and spanning a surface of constant $r$ in the AdS space.
${\rm Vol}(W_5)$ is the two cycle times the volume in AdS bounded
by the surface of constant $r$. The other boundary of the volume, at
$r \to \infty$, gives a contribution $\sim r^{-3}\to 0$ so it does not
contribute.

The supergravity fields obey
\begin{equation}
  \label{eq:151}
  \tilde F_6 \equiv *\tilde F_4
  = F_6 - C_3\wedge H_3 + \frac{m_0}{6}B_2\wedge B_2\wedge B_2.
\end{equation}
Integrating over ${\rm Vol}(W_5)$ we get that the last two terms
vanish, since there are no such background fields with indices in the
non-compact space. We can then write
\begin{equation}
  \label{eq:152}
\int_{{\rm Vol}(W_5)} *\tilde F_4
= \int_{{\rm Vol}(W_5)} F_6.
\end{equation}
The right-hand side is what we want to calculate, while the left-hand side is
proportional to the integration of $\tilde F_4$ over the dual cycle
which we can calculate.

In massive type IIA supergravity we have
\begin{eqnarray}
  \label{eq:153}
  \tilde F_4 &= dC_3 +F_4^{bg}- C_1\wedge H_3 -\frac{m_0}{2} B_2\wedge B_2.
\end{eqnarray}
since the $B_2$ and $C_1$ are only the fluctuations and vanish in the
background, we can replace $\tilde F_4$ in the integral by $F_4$.

In addition to a boundary term we are left with
\begin{equation}
  \label{eq:154}
  \int_{w_2\times w_3}\tilde F_4
  =\int_{w_2\times w_3} F_4^{\rm bg},
\end{equation}
and the WZ term can now be written as
\begin{equation}
  \label{eq:155}
  \mu_4\sqrt2\int_{{\rm Vol}(W_5)} *F_4.
\end{equation}
Now
\begin{eqnarray} F_4&=&e_i\tilde\omega^i=-\({3\over\kappa}\)^\frac13 e_1
(dz^2\wedge d\bar z^2)\wedge (dz^3\wedge d\bar z^3)+\dots, \cr
\ast F_4&=&-4\({3\over\kappa}\)^\frac13 e_1{\gamma_1\over\gamma_2\gamma_3}\Omega_{AdS_4}\wedge
 (du^1\wedge dv^1)+\dots~ \cr
 &=&-2e_i\gamma_i\frac{\gamma_i}{\gamma_1\gamma_2\gamma_3}\(\frac{\sqrt3}{\kappa^2}\)^{1/3}\omega_i,
 \end{eqnarray}
where $\Omega_{AdS_4}$ is the volume form in $AdS$. We find
\begin{eqnarray}
  I_{\rm WZ}&=&\mu_4\sqrt{2}\int_W{\CC_5}
  =-\mu_44\sqrt2\(3\over\kappa\)^\frac13 {a\over 2\kappa^\frac13 3^\frac16}
  {e_1v_1\over v_2v_3}{r^3 \over R_{AdS}^4}\int d^3x
  \cr
  &=&-\mu_43^{-{13\over 12}}2^\frac72 5^\frac14
  {\kappa^\frac13 E^\frac34\over e_1pm_0^\frac14}
  \ {r^3 \over R_{\rm AdS}^3}a\int d^3x
  =-I_{DBI}.
\end{eqnarray}
Thus, the gravitational force due to the DBI term is canceled
exactly by the force from the WZ term, as must be the case for a BPS
configuration.

The analysis is very similar for a more general cycle, and we will not
write it down explicitly here.
\section{Other Domain Walls}
\label{sec:other-domain-walls}

Here we consider D2 and D6-branes in domain wall configurations and
study their supersymmetry equations. We show that a D2-brane can never
be supersymmetric. A D6-brane can classically be supersymmetric,
however due to flux quantization there are generically no such solutions
with integer values for the flux.

\vspace{20pt}
\noindent
{\bf \large C.1 \hspace{5pt} D2-Brane as a Supersymmetric Domain Wall}
\vspace{10pt}

The $\kappa$-symmetry equation (\ref{eq:62}) takes a simple form when
we consider D2-branes, since they are not extended along any compact
dimension. In order for these domain walls to be supersymmetric we
need
\begin{equation}
  \label{eq:110}
  \gamma_{\underline{012}}\eps_-=\eps_+.
\end{equation}
This can be brought to the form
\begin{equation}
  \label{eq:111}
  \gamma_{\underline{012}}\theta_+\otimes b^*\eta_-
  =\theta_-\otimes a^*\eta_-.
\end{equation}
The $AdS_4$ part of the equation is the same as for D4-branes, which
results in the equation
\begin{equation}
  \label{eq:112}
  b^*=\alpha a^*=-{\rm sign}(p)i\frac{b^*}{b}a^*
  \quad\to\quad
  b=-{\rm sign}(p)ia^*,
\end{equation}
which contradicts our supersymmetric condition in the bulk,
$b=-a^*$. Therefore we conclude that D2-branes cannot be
supersymmetric domain walls in this background.

\vspace{20pt}
\noindent
{\bf \large C.2 \hspace{5pt} D6-Brane as a Supersymmetric Domain Wall}
\vspace{10pt}

Consider now a D6-brane which extends as a domain wall in the AdS and
wraps (for instance) the 4-torus spanned by $z_1,z_2$. Its embedding may
be chosen as
\begin{equation}
  \label{eq:113}
  \sigma^1=x^1,\quad \sigma^2=y^1,\qquad
  \sigma^3=x^2,\quad \sigma^4=y^2,
\end{equation}
with the induced metric being
\begin{equation}
  \label{eq:114}
  \gamma_1 (d\sigma^1)^2 +\gamma_1 (d\sigma^2)^2+
  \gamma_2 (d\sigma^3)^2 +\gamma_2 (d\sigma^4)^2,
\end{equation}
and the pullback of $J$ given by
\begin{eqnarray}
  \label{eq:115}
  P[J]=\gamma_1d\sigma^1\wedge d\sigma^2
  +\gamma_2d\sigma^3\wedge d\sigma^4.
\end{eqnarray}

Plugging into the supersymmetry condition (\ref{eq:68}) and taking
$\CF=0$ as for the D4-branes, we have on the right-hand side
\begin{equation}
  \label{eq:116}
  -a^*\alpha\sqrt{\det(P[g]+\CF)}
  d\sigma^1\wedge d\sigma^2\wedge d\sigma^3\wedge d\sigma^4
  =-\sign(p)ib^*
  \gamma_1\gamma_2 d\sigma^1\wedge d\sigma^2\wedge d\sigma^3\wedge d\sigma^4
\end{equation}
while on the left-hand side we have
\begin{align}
  \label{eq:117}
  &\{b^*P[e^{-iJ}])\wedge e^\CF\}_{4}
  =b^* \frac12(-iP[J])\wedge(-iP[J])
  =b^* \frac12(P[J])\wedge(P[J]))
  \cr
  &=b^* \gamma_1\gamma_2 d\sigma^1\wedge
  d\sigma^2\wedge d\sigma^3\wedge d\sigma^4.
\end{align}
Since the first is purely imaginary and the second is real they cannot
be equal, and the domain walls are not supersymmetric. A more generic
embedding will not be able to compensate for the factor of $i$, and so
even the general case is not supersymmetric. However as we have seen
for the D4-branes, adding non trivial $\CF$ can add a relative phase
between the two sides. With $\CF=f_1d\s^1\wedge d\s^2+f_2d\s^3\wedge
d\s^4$ the $\kappa$-symmetry equation becomes
\begin{equation}
  \label{eq:135}
  b^* (f_1-i\g_1)(f_2-i\g_2)
  = -\sign(p)ib^*\sqrt{(\g_1^2+f_1^2)(\g_2^2+f_2^2)}.
\end{equation}
Writing $f-i\g=\sqrt{f^2+\g^2} e^{-\tan^{-1}(\g/f)}$ we get that for
positive $p$
\begin{equation}
  \label{eq:174}
 \tan^{-1}(\g_1/f_1)+\tan^{-1}(\g_2/f_2)=\pi/2
\end{equation}
which has a solution for $f_i>0$. For negative $p$ the right hand side
should be $3\pi/4$ so there are solutions for $f_i>-\g_i$. Classically
such supersymmetric configurations exist, however generically there
are no such configurations consistent with flux quantization.

\end{document}